\documentclass[12pt,preprint]{aastex}

\def\p{\partial}

\def\nab{\mbox{\boldmath $\nabla$}}

\def\rb{\bar{\rho}}
\def\tb{\bar{T}}
\def\sb{\bar{S}}

\def\vph{\hat{v}_{\phi}}

\def\vrr{\tilde{v}_r} 
\def\vtr{\tilde{v}_{\theta}}

\def\vvr{\tilde{v}}
\def\Omh{\hat{\Omega}}

\newcommand{\del}{\mbox{$\nabla$}}

\def\vec#1{\mbox{\boldmath$\displaystyle#1$}}
\newcommand{\dotp}{\vec{\times}}

\begin{document}

\title{Simulations of core convection in rotating A-type stars:  
Differential rotation and overshooting}

\author{Matthew K. Browning, Allan Sacha Brun\altaffilmark{1}, and Juri Toomre}

\affil{JILA and Department of Astrophysical and Planetary Sciences,
University of Colorado, Boulder, CO 80309-0440}

\altaffiltext{1}{New permanenent address: DSM/DAPNIA/Service
d'Astrophysique, CEA Saclay, 91191 Gif-sur-Yvette, France}

\begin{abstract}
We present the results of 3--D simulations of core convection within A-type
stars of 2 solar masses, at a range of rotation rates.  We consider the
inner $30$\% by radius of such stars, thereby encompassing the convective
core and some of the surrounding radiative envelope.  We utilize our
anelastic spherical harmonic (ASH) code, which solves the compressible
Navier-Stokes equations in the anelastic approximation, to examine highly
nonlinear flows that can span multiple scale heights.  The cores of these
stars are found to rotate differentially, with central cylindrical regions
of strikingly slow rotation achieved in our simulations of stars whose
convective Rossby number ($R_{oc}$) is less than unity.  Such differential
rotation results from the redistribution of angular momentum by the
nonlinear convection that strongly senses the overall rotation of the
star. Penetrative convective motions extend into the overlying radiative
zone, yielding a prolate shape (aligned with the rotation axis) to the
central region in which nearly adiabatic stratification is achieved. This
is further surrounded by a region of overshooting motions, the extent of
which is greater at the equator than at the poles, yielding an overall
spherical shape to the domain experiencing at least some convective mixing.
We assess the overshooting achieved as the stability of the radiative
exterior is varied, and the weak circulations that result in that
exterior. The convective plumes serve to excite gravity waves in the
radiative envelope, ranging from localized ripples of many scales to some
remarkable global resonances.

\end{abstract}

\section{INTRODUCTION AND MOTIVATION}

Convection within the cores of massive stars has major impact on their
structure and evolution, yet little is known about the detailed properties
of such convection.  These convective motions, driven by the steep
temperature gradient arising from the vigorous burning of the CNO cycle,
carry outwards a large fraction of the stars' luminosities.  In standard 1--D
stellar models (e.g., Maeder \& Meynet 2000), the effects of such convection
are usually calculated using simple mixing-length prescriptions, but such
approaches involve considerable uncertainties.  Mixing-length theory can provide
only rough estimates of the energy flux carried by the convection, and no
effective estimates of the differential rotation or meridional circulation
generated by the convective flows, which may have important consequences
for mixing and for redistribution of angular momentum within massive stars.

Overshooting from convection zones raises other problems. It has long been
realized that convective motions are unlikely to come to a halt at the
boundary between the convective core and the stable radiative zone above
it. Indeed, upward moving fluid parcels will penetrate into the stable
zone, decelerating and mixing with their surroundings (e.g., Roxburgh
1965).  Such overshooting motions might bring fresh fuel into the core,
thereby prolonging a star's lifetime on the main sequence, and have
noticeable effect on its evolution (e.g., Woo \& Demarque 2001).  Yet
estimating the extent of this overshooting is challenging.

Likewise uncertain is the differential rotation achieved within
convective cores.  In the solar convection zone, Reynolds stresses,
meridional circulations, and viscous forces give rise to a prominent
differential rotation, now being probed extensively by helioseismology
(Gough \& Toomre 1991; Schou et al. 1998).  Explaining that differential
rotation within the sun has been a major challenge for theory and simulation in recent
years, with 3--D spherical shell simulations of turbulent convection now
beginning to make contact with the helioseismic results (e.g., Miesch et
al. 2000; Elliott et al. 2000; Brun \& Toomre 2002). The presence of such
non-uniform rotation deep within more massive stars would have major
consequences for the properties of such stars: differential rotation can
give rise to shear instabilities that stir and mix material, and may serve
to build strong magnetic fields through sustained dynamo action.

Indeed, the generation of magnetic fields within stars by dynamo action
must result from large-scale convection in the stellar plasma interacting
with rotation.  In some stars with convective envelopes, such as the sun,
the building of orderly magnetic fields and cyclic activity is thought to
depend sensitively on highly turbulent convection yielding strong
differential rotation, including a tachocline of shear at the interface
between the bottom of the convection zone and the radiative interior (e.g.,
Thompson et al. 2003).  More generally, correlations between magnetic
fields and rotation have been inferred in main-sequence stars ranging from F5 to M9
(Noyes et al. 1984; Mohanty et al. 2002), with
these fields also thought to be a result of dynamo action driven by
convection, whether occupying an envelope or the full interior.  Yet a
comprehensive understanding of how this occurs remains elusive.  More
massive stars with convective cores are similarly likely to admit magnetic
dynamo action, for they too possess the necessary ingredients: intensely turbulent
convection, a highly conductive medium, and global effects of rotation,
all thought to be crucial for the building of magnetic fields (Charbonneau
\& MacGregor 2001).  Whether the
resulting fields are chaotic, or global and orderly, may be a sensitive
matter.

\subsection{Estimates of Overshooting From Convective Cores}

Some theoretical work has provided constraints on the size of the
convective core and the extent of overshooting.  Roxburgh (1978, 1989,
1998; see also Zahn 1991) showed that an upper limit to the total size
of the convective region can be deduced by considering the integral
\begin{equation}
\int_{0}^{r_{c}}{\left(L_{rad} - L\right)\frac{1}{T^2} \frac{dT}{dr} dr} =
\int_{V}{\frac{\Phi}{T} dV} > 0,
\end{equation}
with $L$ the total nuclear luminosity, $L_{rad}$ the radiative luminosity,
$\Phi$ the viscous dissipation per unit volume, and $r_{c}$ the radius of
the convective core including overshooting.  Since within the convective
core some of the nuclear energy must be carried by convection, $L_{rad} <
L$ there.  Thus if viscous dissipation is neglected, there must be an
overshooting region where $L_{rad} > L$, whose extent can be estimated from
this equation.  For small convective cores, these considerations yield an
overshoot region of size $d \approx 0.18 r_{0}$, where r$_{0}$ is the
radius of the convectively unstable region (Roxburgh 1992).  Rosvick \&
VandenBerg (1998) found that observations of the stellar cluster NGC 6819
were best fit by overshooting that is about half the upper limit
provided by Roxburgh's constraint.

Various attempts have been made to estimate the extent of overshooting from
observations, typically expressed in terms of pressure scale
heights. Meynet, Mermilliod, \& Maeder (1993), for example, found that
best-fit isochrones for a large number of clusters required overshooting
that extends for about $0.2-0.3$ times the pressure scale height.  Perryman
et al. (1998) found evidence for a similar degree of overshooting in Hyades
stars.  In the open cluster M67, whose main-sequence turnoff stars are
thought to be near the low mass end of stars possessing convective cores,
some authors have suggested overshooting that is of order $0.1$ pressure
scale heights (e.g., Maeder \& Meynet 1991; Carraro et al. 1994).

\subsection {Challenges Raised by A-type Stars}

Convective cores are realized for main-sequence stars more massive than
about $1.2$ solar masses, thereby providing many stars in which the effects
of overshooting could be assessed.  However, A-type stars exhibit a variety
of peculiarities that have made them the subject of particularly detailed
study (see Wolff 1983, hereafter W83, for a broad review).  We recall that
A-type stars possess both a convective core as well as multiple shallow
convection zones near the surface.  Some of these stars display strong
abundance anomalies, with greatly enhanced rare-earth element abundances
relative to normal stars (e.g., Kurtz 1990).  Surface abundance anomalies
have also been observed in a variety of other stars (Preston 1974; Gehren
1988; Pinsonneault 1997; Abt 2000).  Radiative diffusive separation,
wherein radiation pressure drives outward some elements while others sink,
is a favored explanation for these abundance features, but requires a very
stable radiative zone in which this may be occurring.  Latour, Toomre, \&
Zahn (1981) showed that the H and He envelope convection zones in A-type stars are
likely linked by penetrative motions that would upset the delicate
quiescence needed for radiative diffusion to work effectively.  This
problem can be avoided if helium gravitationally settles out to greater
depths, in which case the deeper convection zone driven primarily by that
element's second ionization would not exist (e.g., Vauclair, Vauclair, \&
Pamjatnikh 1974).  Such settling is impeded by meridional circulations
within the radiative zone (Michaud et al. 1983), which are thought to
increase in amplitude with increasing stellar rotational velocity. It has
recently been shown (e.g., Richard, Michaud, \& Richer 2001) that in some
stars a further iron convection zone is established below the other two
surface convective regions, which may lead to deeper mixing within
such stars (e.g., Vauclair 2003; Richer, Micaud, \& Turcotte 2000).  While
this basic picture explains many of the observed abundance features, many
of the details are not precisely known.

A subset of the A-type stars also possesses strong surface magnetic
features that appear to persist for many rotation periods (W83).  Such
fields may well be oblique rotators of primordial origin (e.g., Mestel
1999).  However, if dynamo-generated fields within the convective core were
able to rise through the radiative zone, possibly by means of magnetic
buoyancy instabilities, they might also influence the surface fields.
Recent numerical studies (MacGregor \& Cassinelli 2003) have provided
tantalizing indications that it may indeed be possible for strong magnetic
fields generated in the core to rise to the surface.

The A-type stars also hold interest because of the rich set of
observational constraints beginning to be provided by asteroseismological
probing of such stars.  The pulsating Ap stars exhibit high-order nonradial
p-modes, which allow some deductions about stellar radii, temperature, and
magnetic fields (e.g., Matthews 1991; Cunha 2002).

\subsection{Modeling Convection in Full Spherical Domains}

The extensive observations of A-type stars, and the challenges raised, have
encouraged us to consider the core convection influenced by rotation that is
occurring deep within their interiors.  The surface pathologies
of A stars clearly raise many puzzles about their interior dynamics,
thereby lending vibrancy to their study.  We hope that our modeling will
further serve to reveal general dynamical properties of the core convection
that is also occuring in other massive stars.

The major uncertainties associated with core convection -- the extent of
its overshooting into the surrounding radiative envelope, and the
differential rotation and circulations it establishes -- now lead us to
undertake simulations of such convection in full spherical geometries that
permit global connectivity.  We aim to capture much of the essential
physics, admitting highly nonlinear flows that can extend over multiple
scale heights as they mold the dynamical structure of the convective core.
We realize that there may be few observable consequences at the surfaces of
these stars of what may be proceeding dynamically at their
centers. However, the circulations and gravity waves that may be induced by
the core convection may well have implications at the surface, as might magnetic
fields being produced by dynamo action deep within these stars.  All may
depend somewhat sensitively on the rotation rates of the stars.

Advances in supercomputing are now beginning to allow us to examine the
properties of core convection in some detail.  We begin here by considering
3--D hydrodynamic simulations of the inner regions of a 2-solar mass A-type
star.  In subsequent papers, we plan to explore the magnetic dynamo action
that may be realized in such convective cores, and to examine possible
instabilities that might allow the resulting magnetic fields to rise to the
surface.

In Section $2$ we describe our formulation of the problem, and briefly
summarize the computational tools being used.  Section $3$ discusses the
general properties of the nonlinear convective flows realized in the core,
including the transport of energy achieved. The redistribution of angular
momentum by the convection yields prominent differential rotation profiles
that are presented in Section $4$. Section $5$ considers the temporal
evolution of convective patterns within the core, and examines the gravity
waves that are excited within the radiative envelope.  Analysis of the
penetration and overshooting by the convection is discussed in Section $6$.
The meridional circulations induced by the convection, within both the core
and the radiative envelope, are considered in Section $7$.  Section $8$
evaluates the manner in which the convection yields strong differential
rotation.  A summary of our principal findings and their implications is
presented in Section $9$.

\section{POSING THE PROBLEM}

\subsection {Anelastic Equations}

The simulations described here were performed using our Anelastic Spherical
Harmonic (ASH) code.  ASH solves the three-dimensional anelastic equations
of motion in a rotating spherical geometry using a pseudospectral
semi-implicit approach (Clune et al. 1999; Miesch et al. 2000).  These
equations are fully nonlinear in velocity variables and linearized in
thermodynamic variables with respect to a spherically symmetric mean state.
This mean state is taken to have density $\bar{\rho}$, pressure $\bar{P}$,
temperature $\bar{T}$, specific entropy $\bar{S}$; perturbations about this
mean state are written as $\rho$, $P$, $T$, and $S$.  Conservation of mass,
momentum, and energy in this rotating reference frame are therefore written as
\begin{equation}
\nab\cdot(\rb {\bf v}) = 0,
\end{equation}
\vspace{-0.5cm}
\begin{eqnarray}
\rb \left(\frac{\p {\bf v}}{\p t}+({\bf v}\cdot\nab){\bf v}+2{\bf
\Omega_o}\times{\bf v}\right)
 = -\nab P \nonumber \\ + \rho {\bf g} - \nab\cdot\mbox{\boldmath $\cal
D$}-[\nab\bar{P}-\rb{\bf g}],
\end{eqnarray}
\begin{eqnarray}
\rb \tb \frac{\p S}{\p t}&=&\nab\cdot[\kappa_r \rb c_p \nab
(\tb+T)+\kappa \rb \tb \nab (\sb+S)] \\ &-&\rb \tb{\bf v}\cdot\nab
(\sb+S)+2\rb\nu\left[e_{ij}e_{ij}-1/3(\nab\cdot{\bf v})^2\right]
+ \rb \epsilon,
\nonumber
\end{eqnarray}
where $c_p$ is the specific heat at constant pressure, ${\bf
v}=(v_r,v_{\theta},v_{\phi})$ is the local velocity in spherical
geometry in the rotating frame of constant angular velocity ${\bf
\Omega_o}$, ${\bf g}$ is the gravitational acceleration, $\kappa_r$ is the
radiative diffusivity, $\epsilon$ the heating rate per unit mass due to
nuclear energy generation, and ${\bf \cal D}$ is the viscous stress tensor,
with components
\begin{eqnarray}
{\cal D}_{ij}=-2\rb\nu[e_{ij}-1/3(\nab\cdot{\bf v})\delta_{ij}],
\end{eqnarray}
where $e_{ij}$ is the strain rate tensor.  Here $\nu$ and $\kappa$ are
effective
eddy diffusivities for vorticity and entropy.  To close the set of equations,
linearized relations for the thermodynamic fluctuations are taken as
\begin{equation}\label{eos}
\frac{\rho}{\rb}=\frac{P}{\bar{P}}-\frac{T}{\tb}=\frac{P}{\gamma\bar{P}}
-\frac{S}{c_p},
\end{equation}
assuming the ideal gas law
\begin{equation}\label{eqn: gp}
\bar{P}={\cal R} \rb \tb ,
\end{equation}

\noindent where ${\cal R}$ is the gas constant.  The effects of
compressibility on the convection are taken into account by means of the
anelastic approximation, which filters out sound waves that would otherwise
severely limit the time steps allowed by the simulation.

Convection in stellar environments occurs over a large range of scales.
Numerical simulations cannot, with present computing technology, consider
all these scales simultaneously.  We therefore seek to resolve the largest
scales of the nonlinear flow, which we think are likely to be the dominant
players in establishing differential rotation and other mean properties of
the core convection zone.  We do so within a large-eddy simulation (LES)
formulation, which explicitly follows larger scale flows while employing
subgrid-scale (SGS) descriptions for the effects of the unresolved
motions.  Here, those unresolved motions are treated as enhancements to the
viscocity and thermal diffusivity ($\nu$ and $\kappa$), which are thus
effective eddy viscosities and diffusivities.  For simplicity, we have
taken these to be functions of radius alone, and to scale as the inverse of
the square root of the mean density.  We emphasize that currently tractable simulations are
still many orders of magnitude away in parameter space from the highly
turbulent conditions likely to be found in real stellar convection zones.
These large-eddy simulations should therefore be viewed only as indicators
of the properties of the real flows.  We are encouraged, however, by the
success that similar simulations (e.g., Miesch et al. 2000; Elliott et
al. 2000; Brun \& Toomre 2002) have enjoyed in matching the detailed
observational constraints for the differential rotation within the solar
convection zone provided by helioseismology.

\subsection {Computational Approach}

Thermodynamic variables within ASH are expanded in spherical harmonics
$Y^m_{\ell}(\theta,\phi)$ in the horizontal directions and in Chebyshev
polynomials $T_n (r)$ in the radial.  Spatial resolution is thus uniform
everywhere on a sphere when a complete set of spherical harmonics of degree
$\ell$ is used, retaining all azimuthal orders $m$.  We truncate our
expansion at degree $\ell=\ell_{max}$, which is related to the number of
latitudinal mesh points $N_{\theta}$ (here $\ell_{max}=(2N_{\theta}-1)/3$),
take $N_{\phi}=2 N_{\theta}$ latitudinal mesh points, and utilize $N_r$
collocation points for the projection onto the Chebyshev polynomials.  We
employ a stacked Chebyshev representation, wherein the computational domain
is split into two regions and separate Chebyshev expansions performed for
each.  The interface between these two regions, here taken to be the
approximate boundary between the convective and radiative zones, is thus
treated with greater resolution in order to capture elements of the
penetrative convection occurring there.  We have taken $N_r=49+33=82$ and
$\ell_{max}=85$ in this study for the cases considered here, and have used
$\ell_{max}=170$ for some companion simulations to verify that the spectral
resolution was adequate.  An implicit, second-order Crank-Nicholson
scheme is used in determining the time evolution of the linear terms,
whereas an explicit second-order Adams-Bashforth scheme is employed for the
advective and Coriolis terms.  The ASH code has been optimized to run
efficiently on massively parallel supercomputers such as the IBM SP-3 and
the Compaq TCS-1, and has demonstrated excellent scalability on such machines.

\subsection {Formulation of the Model}

The model considered here is intended to be a simplified description of the
inner 30\% by radius of a main sequence A-type star of 2 solar masses,
consisting of the convective core (approximately the inner 15\% of the
star) and a portion of the overlying radiative zone.  Contact is made with
a 1--D stellar model (at an age of 500 Myr) for the initial conditions,
adopting realistic values for the radiative opacity, density, and
temperature.  This 1--D model was obtained using the CESAM stellar
evolution code (Morel 1997, Brun et al. 2002), which employs the OPAL
equation of state and opacities (Iglesias \& Rogers 1996; Rogers, Swenson,
\& Iglesias 1996) and the nuclear cross sections of Adelberger et
al. (1998) in describing the microscopic properties of the stellar plasma.
Convection is computed within CESAM using a classical mixing length
treatment calibrated on solar models, with some convective overshooting
(typically $r=0.25 H_p$) taken into account.  Our simulations were
initialized using the radial profiles of gravity $g$, radiative diffusivity
$\kappa_{rad}$, mean density $\bar{\rho}$, and mean entropy gradient
$d\bar{S}/dr$ as the starting points for an iterative Newton-Raphson
solution for the hydrostatic balance and for the gradients of the
thermodynamic variables.  The mean temperature $\bar{T}$ is then deduced
from equation (7).  This technique yields mean profiles in reasonable
agreement with the 1--D stellar model (Fig. $1$).  The nuclear source of
energy $\epsilon$ in our simulations is also deduced from the stellar
model.  For computational convenience, we take $\epsilon =
\epsilon_0 \bar{T}^8$, determining the constant $\epsilon_0$ by requiring
that the integrated luminosity match the 1--D model's surface luminosity.  

We have softened the steep entropy gradient contrast encountered in going
from the convective core to the surrounding radiative zone, which would
otherwise favor the driving of small-scale, high-frequency internal gravity
waves that are presently unresolvable given reasonable computational
resources.  This lessening of the `stiffness' of the system has impact on
the extent to which convective motions may overshoot into the radiative
region.  We have excluded the inner 2\% of the star from our computational
domain for numerical reasons, namely that the coordinate systems employed
in ASH possess both coordinate singularities at $r=0$ and decreasing
meshsize (and hence highly limited time steps) with decreasing radius. It
is difficult to gauge with certainty the effects of excluding this central
region, which might in principle project both a Taylor column aligned with
the rotation axis (e.g., Pedlosky 1987) into the surrounding fluid, as well
as give rise to boundary layers.  However we have seen no evidence that
spurious physical responses have been generated by its omission.  In trial
simulations with both smaller and larger excluded central regions (ranging
from 0.3\% to 4\% of the stellar radius), the properties of the developed
mean flows were very similar to those described here.

As boundary conditions on the spherical domain, we utilize impenetrable and
stress-free conditions for the velocity field and constant entropy gradient (i.e
constant emergent flux) both at the inner and outer boundaries.  As initial
conditions, some simulations have been started from quiescent conditions of
uniform rotation whose radial stratification is given by the 1--D stellar
model; others used evolved states from prior solutions to initiate
simulations with reduced diffusivities.  

The main parameters of our simulations are summarized in Table 1.  In
brief, we have modeled the inner regions of 2-solar mass A-type stars at a
range of rotation rates, with nonlinear flows of varying complexity
achieved by modifying the effective eddy viscosities and diffusivities
$\nu$ and $\kappa$.  Several values of $\nu$ and $\kappa$ are utilized to
achieve both laminar and more complex convective flows in some companion
cases, all at a Prandtl number $P_r = \nu/\kappa = 0.25$.  We consider rotation
rates from one-tenth to four times the solar rate of {\bf $\Omega_{o}$}$=
2.6\times 10^{-6}$ $s^{-1}$ (or $414$ nHz), though concentrating primarily
on models at the solar rotation rate.  Three different values of the
maximum entropy gradient $d\sb/dr$ in the radiative region were taken, thus
effectively varying the stiffness of the boundary between that region and
the convective core, primarily as a way to study the effects on
overshooting motions induced by softening that boundary.  We have chosen
only representative simulations for summary in this paper (and in Table 1)
out of a larger total set of computations.

Figure $1$ shows the mean radial profiles for temperature, density, and
pressure attained in a representative simulation (case C), together with
profiles of the same variables in a typical 1--D stellar structure model.
The evolved simulation possesses somewhat less steep gradients for these
variables than does the 1--D stellar model, but the functional form of
these quantities is largely unchanged by our detailed treatment of the
convection.


\section{NATURE OF CORE CONVECTION} 

The flows in all the cases studied are highly time dependent, with complex
and intermittent features emerging as the simulations evolve.  The
convection is characterized by large-scale flows that sweep through much of
the unstable core, coupling widely separated sites.  Thus influences of the
convective motions are decidely nonlocal.  Such global motions can result
because of the topologically connected nature of full spheres.  The
addition of rotation to these spherical domains admits Coriolis forces and
resulting velocity correlations (Reynolds stresses) that yield some
surprising effects.  In particular, we find that the cores exhibit
prominent differential rotation both in radius and latitude, involving in
most of our simulations a central column of slow rotation aligned with the
rotation axis.  Further, the extent of penetration of the convection into
the surrounding stable radiative envelope varies from equator to pole,
yielding a prolate shape to the central region of nearly adiabatic
stratification.

The rich time dependence typical of core convection implies that extended
simulation runs must be conducted in order to sample the dynamics.
Attaining a statistically equilibrated state in the dynamics typically
requires evolving the flows for about $1500$ days of physical time, or
roughly 60 rotational periods at the nominal solar rate. Deductions about
the mean flows accompanying the convection, such as differential rotation
and meridional circulation, or of the mean heat transport achieved by the
convection, require averaging the flows over lengthy periods of time,
typically greater than 100 days in the late stages of a simulation. We
begin by describing some of the general attributes of the convection
realized in our simulations.

\subsection{Topology and Scales of Convective Flows}

Core convection clearly involves global-scale overturning motions that
extend over much of the unstable domain.  This is illustrated in Figure $2$
which presents both a 3--D rendering of the instantaneous radial velocity
near the outer surface of the convective core studied in case E, and an
equatorial cut at the same instant through that core domain.  Specific
regions of upflow and downflow occupy large fractions of the irregular
surface (Fig. $2a$), and extend in radius over much of that unstable volume
(Fig. $2b$).  Evidently there is global connectivity in the flows over well
separated locations, spanning wide ranges in longitude and latitude in both
hemispheres.  The equatorial cut reveals that the flow patterns
involve a prominent azimuthal wavenumber $m=2$ component, though many other
wavenumbers are certainly also involved in describing these overturning
motions.  Little asymmetry is apparent between upflows and downflows:
radially inward and outward motions occur on roughly the same spatial
scales, and possess comparable flow amplitudes (see also Table 2).  This
appears to be a consequence of the relatively modest radial density
contrast (about 2.5) present across the unstable core.  We do not here
obtain the fast narrow downflows and much broader upflows realized in far
more compressible configurations representative of the solar convection
zone (Brummell et al. 1998, Brun \& Toomre 2002).

The convectively unstable region that results is mildly prolate in shape (Fig. $2a$), having 
greater spatial extent near the poles than near the equator.  This non-spheroidal 
shape (cf. \S6.1) for the extent of the vigorous convection is a striking feature 
shared by all our simulations.  The rumpled appearance of $v_r$ in this 3--D volume
rendering comes from emphasizing the large-scale penetrative structure of such velocity 
fields. In studying time sequences of such renderings, the interface between the
well-mixed interior and the radiative exterior looks much like a throbbing heart,
as penetrative motions wax and wane and propagate laterally, launching gravity
waves in the process (cf. \S5.2).  Analysis of the time evolution of
equatorial cuts (Fig. $2b$) reveals that some flows dive through the
central regions of the core to emerge on the other side, with such patterns
exhibiting distinctive retrograde motion.

\subsection {Velocity and Thermal Patterns}

We present in Figure $3$ global mappings for cases B and C of both the radial 
velocities $v_r$ and temperature fluctuations $T$ (relative to the average of 
temperature on a spherical surface at a given radius).  These Mollweide projections 
for the fields are shown both well within the convective interior (at $r=0.10R$, 
lower panels) and at the transition into the stable exterior ($r=0.16 R$, 
upper panels).  These mappings reveal that the radial velocity patterns in any given
snapshot have a noticeable alignment with the rotation axis.  (Such projections 
exhibit meridian lines as increasingly curved arcs away from the central meridian, 
which appears as a
vertical line; here the lines of constant latitude are indeed parallel.)  The more 
laminar case B (Figs. $3a-d$, left panels) possesses structures that are largely 
symmetric about the equator, whereas the lower diffusivities ($\nu$ and $\kappa$) 
in case C (Figs. $3e-h$, right) lead to flows with broken symmetries in the two 
hemispheres as more intricate flows are realized.  In both cases the flow structures 
at high latitudes are more isotropic and of smaller scale, and somewhat faster 
flows are realized for case C (see also Table $2$).

Radial motions in these simulations are intimately linked to fluctuations
in temperature.  These buoyantly driven flows should typically involve positive $T$
in regions of ascending motion within the unstable core, and negative $T$
in descending flows there.  Such a positive correlation between $v_r$ and $T$
is largely realized in the lower panels in Figure $3$ for both cases
showing the $r=0.10R$ depth where the convection is vigorous.  The less
complex flows in case B exhibit such correlation readily, whereas in case C
there are also sites where the sense of the correlation is reversed.  The
nature of the correlations changes as motions penetrate into the radiative
surroundings. The buoyancy braking that is expected within this region of
stable stratification, as at the $r=0.16R$ depth (upper panels, Fig. $3$),
arises from the radial velocities and temperature fluctuations becoming
anti-correlated.  The sense of $T$ in both regions is partly controlled by
the source term in equation (4) that involves the product of $v_r$ with the
radial mean entropy gradient.  When the latter gradient changes sign in
going from an unstable stratification (superadiabatic in the convective
core) to stable (subadiabatic) stratification as flows extend upward into
the radiative envelope, so does the sense of $T$.  This correspondence
between radial velocities and temperature fluctuations is generally true
only in the large: on finer scales, secondary instabilities can drive
motions that locally violate these principles.  Similarly, the nonlinear
advection of temperature fluctuations by motions that twist and turn can
lead to unusual combinations of $v_r$ and $T$ at some sites.  The
anti-correlation of $v_r$ and $T$ in the region of penetration is clearest
in case $B$ involving the simpler flow fields.  Because the
thermal diffusivity is larger than the viscosity in these simulations
(i.e. the Prandtl number $P_r$ is less than unity, namely 0.25), thermal
structures have a somewhat larger physical scale than do associated
features in the radial velocity.

The temperature fluctuations at both depths in Figure $3$ have amplitudes
of only a few K (compared to the mean temperatures of millions of K found
at such depths), which combined with radial velocities of order 50 m
s$^{-1}$ in the bulk of the convective interior and the high heat capacity
of the gas, are sufficient to carry outward a good fraction of the emerging
flux (cf. \S3.3).  The radial velocities have plummeted to much smaller
values just beyond the edge of the convection zone where the motions are
beginning to penetrate upward into the radiative zone.

The convective fields in these simulations can be assessed using
probability distribution functions (pdfs).  Figure $4$ shows pdfs for the
radial velocity and temperature fluctuations on a spherical surface within
the convective core ($r=0.10R$) and the region of overshooting ($r=0.16R$)
of case C, exhibiting both instantaneous pdfs and time-averaged ones.  The
distributions of radial velocities are largely symmetric Gaussian functions
at both depths. The detailed features in the instantaneous pdfs change as
the flows evolve.  The instantaneous temperature field distributions
(Fig. $4d,h$) possess spikes, indicating the presence of large-scale
coherent temperature structure; at other instants, the spikes may be
replaced by broader plateaus.  The time-averaged pdfs for T are
nonsymmetric, with broad tails extending to positive values in the region
of overshooting and to negative ones in the core.  These tails serve to
compensate for the off-center most probable values that are positive in the
core and negative in the overshooting region. Such complexities in the pdfs
are realized in all of our simulations, and thus are at some variance with
simpler relations assumed in mixing-length treatments of convection.

\subsection {Energy Transport}

The release of energy by CNO-cyle nuclear burning in the core leads to
a progressively increasing local luminosity with radius within the
inner parts of our domain.  The energy produced in such a distributed fashion 
gets carried outward primarily by radiative diffusion and mechanical transport of
heat (or enthalpy) by the convection. 
To assess the role convection plays in the overall energy balance, 
we recall that the rate at which convection transports energy in the radial 
direction is given by the enthalpy flux
\begin{equation}
  F_{e}(r) =  \rb c_{p} \overline{v_{r} T},
\end{equation}
involving a product of the radial velocity with fluctuations in the
heat content of the fluid, proportional to $T$, and averaged over a
spherical surface and in time (indicated by the `overbar').  
We may compare $F_{e}$ with the flux of energy carried by radiation,
\begin{equation}
  F_{r}(r) = - \kappa_{r} \rb c_{p} (\frac{d\tb}{d r} + \overline{\frac{d T}{d r}}) ,
\end{equation}
with the latter term arising from the mean of radial gradients of temperature 
fluctuations, which are here negligible. 
The enthalpy and radiative fluxes largely suffice to carry outwards the energy 
produced by nuclear reactions within the core, yielding a total flux
\begin{equation}
  F_{t} = \frac{L(r)}{4 \pi r^2},
\end{equation}
where $L$ is the luminosity of the star that increases with
radius until the surface luminosity $L_*$ is attained. (The kinetic energy flux, the 
viscous flux, and the flux carried by unresolved subgrid motions are all small 
compared to $F_{e}$ and $F_{r}$.)

Figure $5$ displays the variation with radius of $F_r$ and $F_e$, expressed
as luminosities, for case C.  The enthalpy flux is maximized near the
middle of the convective core (at $r=0.08R$), serving to carry about 57\%
of the local stellar luminosity $L$ there, with the remainder transported
by radiation.  Since the core convection establishes a nearly adiabatic
stratification (with $\nabla - \nabla_{ad} \sim 10^{-7}$), the associated
temperature gradient that arises serves to specify a radiative flux $F_r$
that increases steadily with radius.  Such a growing $F_r$ forces $F_e$ to
decrease in the upper half of the unstable core. Beyond the convectively
unstable region, at $\bar{r}_{c} \sim 0.14 R$ (cf. \S6.2), the enthalpy
flux becomes negative, resulting from the anti-correlation of radial
velocity and temperature fluctuations as the motions are braked.  The
radiative flux there must compensate for that inward-directed enthalpy
flux.  Our simulations have not been evolved for a sufficient amount of
time to allow this compensation to occur fully, since the relevant thermal
relaxation time is long compared to other timescales in the
problem. Thus the total luminosity in Figure $5$ displays a small dip at
the interface between the convective and radiative zones, which in real
stars (or a fully evolved simulation) would be absent.  

The correlations of radial velocity and fluctuating temperature that yield
an outward-directed enthalpy flux are achieved in a complicated manner
within the bulk of the convective interior.  In mapping these variables at
one instant in time, it is evident that a number of sites yield an
unfavorable correlation, whereas in the main the requisite positive
correlation is realized.

The nearly adiabatic stratification achieved by the convection in the bulk of 
the core is much as in 1--D stellar models employing standard mixing-length treatments.  
However, our self-consistent 3--D analysis of convective dynamics and its transports 
also provides estimates of the stratification in the region of overshooting, which 
are far more difficult to assess using mixing-length approaches.


\section {RESULTING DIFFERENTIAL ROTATION}

Convection occurring within deep spherical domains should be strongly
influenced by the rotation of the star when the convective overturning time
is of the order or greater than the rotation period.  This implies a
convective Rossby number $R_{oc}$ (cf. Table 1) which is of order unity or
smaller.  The global-scale convection in our models has an overturning
period of about one month, comparable to the rotation period of a star
rotating at the solar rate.  Thus we expect the convection influenced by Coriolis
forces to possess correlations between the components of its velocity field.
The resulting Reynolds stresses can serve to continuously redistribute
angular momentum, leading to angular velocities within the core that are
far from uniform rotation.  Further, the connected topology of the
convective core appears to play an important role in the prominent
differential rotation profiles that are realized in our simulations.

\subsection {Central Column of Slow or Rapid Rotation}

We find that strong angular velocity contrasts are established within the
convective cores in all our simulations.  For model stars that rotate at
least as rapidly as the sun, these involve roughly cylindrical
central columns of decidedly slow rotation (retrograde relative to the
reference frame), accompanied by equatorial regions of somewhat enhanced
rotational velocity (prograde).  As shown in Figure $6$ (left panels) for
the three cases B, C and C4, the time-averaged mean longitudinal
velocity $\vph$ exhibits some variations along that central column, with
those fluctuations becoming more pronounced as the complexity of the
convection increases in going from case B to C (cf. Fig. $3$).  There is
likewise greater asymmetry in $\vph$ between the two hemispheres
(delineated by the equator) in these cases.  Such symmetry breaking may be
anticipated as the convection becomes increasingly nonlinear. The
convection itself is not symmetric about the equator, and thus mean flows
associated with it, including the differential rotation, are
likely to exhibit differences in the two hemispheres.  The columns of slow
rotation extend slightly into the radiative envelopes at high latitudes, in
keeping with the prolate shape of the convective core.  The presence of the
central slow cylinder of rotation does not appear to result from the small
inner sphere omitted from our computational domain, for we have found
similar columns of slowness for several different sizes of central
spheres.  Figure $6$ (right panels) shows the variation of angular velocity
$\Omh$ with radius in three latitudinal cuts, providing another assessment
of the slowness of the central column and of the speeding up of the
equatorial region.

These central columns of slow rotation are largely unexpected, and
constitute a striking finding of these simulations.  A general result from
previous studies of convection in deep shells under strong rotational
constraints (with $R_{oc}$ small) is that equatorial regions of fast
rotation are obtained (e.g., Gilman 1979; Miesch et al.  2000; Brun \&
Toomre 2002).  The conservation of angular momentum requires that other
regions must be slowed down, and this often appears as higher latitude
regions of slowness.  The analog in our spherical domain may be the central
column of particularly retrograde $\vph$, for this could compensate for the
equatorial speeding up (prograde $\vph$) of regions with a large moment
arm.  The continuous redistribution of angular momentum by the convection,
involving variously effects of Reynolds stresses, meridional circulations
and viscous stresses, will be examined in \S8 as we seek to understand the
balances needed to account for such central columns of slowness.

We have also considered a more slowly rotating case F, with a rotation rate
of $\Omega_o/10$, to examine whether the central column of slowness is
still achieved under a weaker rotational influence (with $R_{oc}$ large).
The prior studies of convection in shells (Gilman 1978) indicated that
under such conditions the equatorial regions may exhibit relatively slow
rotation due to the modified transport of angular momentum.  If this
property of shell convection were to carry over to full spherical domains,
then it might follow that the overall conservation of angular momentum
could yield central regions of fast rotation to compensate for the
equatorial slowness.  Figure $7$ shows that for case F this is indeed
realized.  The central region of relatively fast rotation is decidedly less
columnar than those of slow rotation in Figure $6$, probably because the
Taylor-Proudman constraint is weakened.  We thus suspect that central
columns of slowness may be ubiquitous for stars that rotate sufficiently
rapidly that $R_{oc}$ is less than unity.  In the remainder of this paper,
we concentrate on our simulations that satisfy this criterion (all but case
F), since real A-type stars are generally quite rapidly rotating.

\subsection {Contrasts in Angular Velocity}

We now turn to the time-averaged angular velocity $\Omh$ profiles associated with the
differential rotation in our convective cores with $R_{oc}$ small, some
examples of which are shown as radial cuts at constant latitudes in the
right-hand panels of Figure $6$ that accompany the views of $\vph$.  The
$\Omh$ profiles in Figure $6$ emphasize that most of the radiative
envelope rotates at the nominal rate of the reference frame.  The
variations in $\Omh$ are largely confined to the convective core, though
within the overshoot region there are still some weak gradients in $\Omh$
evident.  In the simulations of semi-turbulent convection rotating at the
solar rate (cases A, C and E, with identical viscosities but increasing
`stiffness' in the stratification of the radiative exterior), we obtain
angular velocity contrasts $\Delta \Omh /
\Omega_o$ (from the equator to 60\degr) of order 60\% (see also Table 2). These contrasts appear to be
insensitive to the stratification of the radiative zone considered here;
likewise the character of the differential rotation throughout the
convective core is similar in these three cases. In more laminar
simulations at the same rotation rate (cases B and D, involving greater
viscosities than their companion cases C and E), though the angular
velocity contrast is lessened to about 25\%, a central cylinder of slow
rotation is still evident.  At a four-fold higher rotation rate considered
as case C4, $\Delta \Omega$ increases by about a factor of two, yielding a
$\Delta \Omega/ \Omega_o$ that is reduced to about 33\%; a central column of
slowness is retained.  As we increased the rotation rate in going to case
C4 from case C, we sought to maintain roughly the same degree of
supercriticality in the convection, which required decreasing the effective
viscosities to compensate for the stabilizing effects of rotation (e.g.,
Chandrasekhar 1961). Without such decreases, the angular velocity contrast
was lessened to about 22\% in the resulting more laminar flow.

Another assessment of the resulting differential rotation in these
simulations is provided by examining the kinetic energies associated
variously with the convection (CKE), meridional circulation (MCKE), and
differential rotation (DRKE), together with their sum (KE), all relative to
the rotating frame.  As detailed in Table 2, the CKE is reasonably
invariant across all our simulations rotating at the solar rate.  In going
from the more laminar flows of cases B and D to the semi-turbulent flows in
cases C and E, the stronger differential rotation that results is reflected
in the greater DRKE, and thus also in the total kinetic energy.  The MCKE
is likewise increased, but remains a small fraction of KE.  Thus the more
complex flows more effectively transfer energy -- from the vast reservoirs
of energy associated with the rotation of the frame and with the internal
energy of the star -- into KE, which nonetheless remains a small fraction
of the total energy in these simulations.

From our limited sampling of rotation rates, we cannot reliably estimate
how $\Delta \Omega / \Omega$ is likely to vary with $\Omega$ at the much
higher rates typical of many A-type stars.  However, we
suspect that the strong differential rotation obtained in our simulations
is likely to be realized even for core convection within more rapidly
rotating stars with much stronger rotational constraints.


\section {EVOLVING CONVECTION AND WAVE EXCITATION}

The inherently time-dependent convection involves substantial pattern
evolution and propagation within the bulk of the convective core, and the
excitation of gravity waves in the surrounding radiative envelope. Some
large features in the convection appear, persist, and are sheared and
advected by the mean flows that they drive. The radiative zone contains the
rippling signature of unremitting gravity waves excited by the convection
below, accompanied by weak circulations.  We now consider such time
variability.

\subsection {Propagation and Shearing of Convective Structures}

The flows in these simulations evolve in a complicated manner.  This may be
seen in Figure $8$, which examines the radial velocity patterns within the
convective core (at $r=0.10 R$) in case C at three instants each separated
by seven days in time.  Large scale features are recognizable in all three
images, but show evolution and propagation.  Some orderly downflow
structures within the convection persist for long periods of time.  Yet
they also clearly evolve over relatively short time intervals: major
downflow lanes present in Figure $8a$, for instance, are recognizable in
Figure $8c$, but have changed in shape and intensity. They do not evolve in
isolation: patterns merge, are sheared, and cleave into smaller patterns,
as they interact with surrounding flows. Such features are not simply
advected by the mean motion of the region in which they are embedded, but
can additionally propagate relative to the differential rotation that they
establish.  The rate and sense of pattern propagation is most readily seen
by turning to time-longitude mappings, such as in sampling the radial
velocities at 65\degr$\ $ north (Fig. $8d$) and at the equator (Fig. $8e$)
over nearly 200 days.  In these mappings, persistent strong downflow lanes
are revealed as dark bands tilted to the right in Figure $8e$, and sheared
to the left in Figure $8d$.  Such tilts reveal that patterns at this radius
and near the equator propagate in a prograde sense, taking about 300 days
to complete one revolution, and at the higher latitude their propagation is
retrograde with a period of about 60 days.  These mappings also emphasize
that the convection possesses larger scales at the equator than at high
latitudes, with a distinctive change in the sense of pattern propagation
relative to the frame at mid-latitudes.

\subsection {Internal Gravity Wave Responses}

The vigorous core convection extends upward into the stable stratification
that bounds it, with such pummeling serving to excite internal gravity
waves within the radiative envelope.  Such wave responses are shown vividly
in the upper panels of Figure $9$, which displays the radial velocities at
multiple depths in cases E and C4.  These can appear both as localized
ripples that are one signature of the waves (Fig. $9a$), arising either
from distinctive events or from the interference of many waves, or as
striking large-scale global resonances (Fig. $9d$) that persist (once
established) for as many rotation periods as we continued the
simulations. The latter response may be more easily realized in our
simulations, with their artificially imposed outer boundary conditions that
reflect gravity waves, than in real stars where the radiative envelope is
extensive and probably has no simple reflecting surfaces.  However, they
may also mimic gravity waves of moderate radial order $n$ (such as $n=8$ or
$10$) in a real star, though the stable stratifications in such stars are
much stiffer than the ones we consider here.

The generation of gravity waves by the upward-directed plumes is fully
expected.  Waves with frequencies less than or equal to the Brunt-Vaisala
frequency $N$ (proportional to the square root of the local entropy
gradient $d \sb/dr$ of the stable stratification) are able to propagate in
the radiative outer envelope and may be excited in abundance by the
time-varying convection, as in case E at $r=0.26 R$ (Fig. $9a$).  This also
implies that the spectrum will shift as the stiffness of the surrounding
envelope is varied.  More surprising is the large-scale global resonance of
an internal gravity mode (of spherical harmonic degree $\ell=3$ and
azimuthal order $m=2$) that is readily achieved within the more rapidly
rotating case C4, and likewise within some of its more laminar relatives.
This may come about because the rotating convection is likely to possess a
range of inertial oscillations with frequencies less than 2$\Omega_o$
(Greenspan 1969), which accompany the broad band of lower frequencies
present in the time-dependent convection. When there is a suitable overlap
between the temporal spectrum of the rotating convection and the admissible
internal gravity wave frequencies that depend on $\ell$ and radial order
$n$ (e.g., Unno et al. 1979; Gough 1993; Dintrans, Rieutord, \& Valdettaro
1999), a prominent wave response may result.  In our models $N$ varies
somewhat with radius; for case C4 its representative value is $1.1 \times
10^{-5}$ Hz, whereas $\Omega_o$ is $1.04 \times 10^{-5}$ Hz.  It appears
that a global gravity wave (with low $\ell$) has been selected that
propagates retrograde relative to the frame (with a period of about $40$
days for a complete revolution), within a region that exhibits almost no
differential rotation. In addition to the global mode, other gravity waves
appear as ripples near the equator in Figure $9d$.


\section {CONVECTIVE PENETRATION AND RADIATIVE ENVELOPE}

Convective motions do not stop abruptly at the interface between the
convective core and the stable radiative envelope.  Rather, outward moving
parcels of fluid possess momentum, and therefore must travel some distance
into the stable radiative zone before they come to a halt.  There are two
effects.  Firstly, if convective motions are sufficiently vigorous, and
possess a large enough filling factor, they can create a nearly-adiabatic
stratification within a portion of the overlying radiative zone that in a
1--D stellar structure model would be stably stratified.  This may be
termed a region of \emph{penetration}, as discussed in Zahn (1991),
Hurlburt et al. (1994), and Brummell, Clune, \& Toomre (2002).  Secondly, there will
be a further region of stable stratification into which convective motions
extend, but whose entropy gradient they fail to modify significantly.
Within such a region of \emph{overshooting}, outward moving parcels of
fluid are braked until they come to a halt.  Throughout the domain of
penetrative convection, and the extended region of overshooting, the fluid
motions should be effective in mixing the chemical composition.

\subsection {Shape of Convective Core}

In the 1--D stellar models used for our initial conditions, the region
of unstable stratification is of course spherical.  In the evolved
simulations, penetrative convective motions have yielded a nearly
adiabatically stratified core region (weakly superadibatic) that is now
prolate in shape and aligned with the rotation axis.  The radius
$r_{c}$ of this core region thus varies with latitude, as sketched in
Figure $10$.  The simulations further reveal that the overshooting
motions extend outward to a radius $r_{o}$ that is largely independent
of latitude, and thus roughly spherical in shape, as also sketched.

Such geometries for the penetrative core and the further region of
overshooting are shown in Figure $11$ for case C.  These regions are
delineated by turning to the time- and longitude-averaged enthalpy flux,
given by
\begin{equation}
  \hat{F_e}(r,\theta) =  \rb c_{p} <v_{r} T>_{t,\phi},
\end{equation}
where the brackets and their subscripts indicate the averaging.  Within the
convective core that includes the region of penetration, the enthalpy flux
is typically positive, owing to the positive correlations of radial
velocity and temperature fluctuations that are largely realized there
(cf. \S 3.2).  As the motions extend outward into the subadiabatic
stratification of the region of overshoot, the enthalpy flux becomes
negative since the temperature fluctuations typically flip their sign
almost immediately, becoming in the main anti-correlated with the radial
velocity.  Thus the inner boundary of the region of overshoot is usefully
also revealed by the radius where $\hat{F_e}$ changes sign.

The outer boundary $r_o$ of that overshooting region is taken to be the
radius at which the enthalpy flux has attained an amplitude of 10\% of its
most negative value.  We introduce this definition because small negative
values of the enthalpy flux persist wherever any radial motions
anti-correlated with temperature fluctuations exist, even if these radial
motions are very weak. Our intent is to identify regions where overshooting
motions are reasonably vigorous, recognizing that such a constraint on the
amplitude of the enthalpy flux is somewhat arbitrary. For case C, $r_o$ has
a value of about $0.158 R$, independent of latitude, whereas $r_c$ ranges
from about $0.131 R$ at the equator to $0.148 R$ at the poles (Fig. $11$).
The values for other cases are presented in Table 2. Since in all our cases
the inner boundary $r_c$ of the overshoot region is a function of latitude
and the outer boundary $r_o$ is largely not, the extent of the overshooting
$d_o$ varies with latitude, possessing the greatest spatial extent near the
equator.  Figure $11$ also shows that the enthalpy flux $\hat{F_e}$ has
prominent variations with latitude and radius within the convective core,
with a greater flux at mid and high latitudes than at the equator.  As the
convection evolves, the time-averaged $\hat{F_e}$ as determined over
different epochs shows some changes in its structure and its symmetries
with respect to the equator.

The prolate geometry of the convective core may result from the effects of
rotation yielding convective flows that typically involve greater radial
velocities at high latitudes than at the equator.  Figure $12$ shows the rms
radial velocity in case C within both the convective core ($r=0.10 R$) and
the region of overshooting ($r=0.15 R$). It is evident that radial
velocities are generally lowest near the equator and highest near the
poles.  Outward-directed motions at low latitudes are turned by the
Coriolis forces, thus decreasing their radial velocity component;
radial motions that are closely aligned with the rotation axis, such as at
high latitudes, are deflected less, and thus retain most of their radial
velocity.

\subsection {Stable Stratification and Overshooting}

As convective motions enter the stable stratification of the radiative
envelope, they are buoyantly braked (cf. \S3.2).  The amount of braking
they experience is controlled by the magnitude of the stable entropy
gradient.  By varying that gradient in our simulations, we have thus
effectively altered the ability of flows to overshoot into the radiative
envelope: a larger $d\sb/dr$ implies stronger buoyancy braking, and thus a
smaller extent to the overshooting region.  A larger $d\sb/dr$ also leads
to a smaller extent to the penetration as indicated by $r_c$. Both these
effects are apparent in Figure $13$, which displays $F_{e}(r)$ at several
latitudes, for three simulations of varying $d\sb/dr$.  With increasing
stiffness comes decreasing $r_o$ and $r_c$.

For our stiffest and most complex case E, the latitude-averaged extent of
overshoot, $\bar{d_{o}}$, is $1.906 \times 10^{9}$ cm, or $ 0.208 H_{p}$,
where $H_{p}$ is the local pressure scale height averaged over the region
of overshoot.  In the less stiff case C, a larger $\bar{d_{o}}$ is realized; such
measures are quoted for all cases in Table 2.  Increasingly complex flows,
as achieved in our simulations by lowering the viscosity in going from case
B to case C (or from case D to case E, at a different stiffness), yield
smaller $\bar{d_o}$ than do their more laminar relatives.  This seems at first
paradoxical, given the larger rms velocities associated with the increased
complexity of the convection. However, this trend is largely in keeping
with nonlinear modeling of 3--D penetrative compressible convection in
localized planar domains (cf. Brummell et al. 2002), which
showed that a smaller filling factor for the plumes as realized in the more
turbulent convection translates into a smaller extent to the overshooting.

Increasing $\Omega_o$ without altering the viscosity leads to more laminar
convection, due to the stabilizing effects of rotation.  This in turn
yields a greater overshooting. In devising case C4 with its fourfold
greater rotation rate, we reduced the viscosity relative to case C in order
to achieve comparable complexity in the convection as measured by the
Reynolds numbers.  In case C4, $\bar{d_o}$ is greater than that in the
slower rotating case C, suggesting that increasing the rotation rate
enhances somewhat the overshooting.


\section {INDUCED MERIDIONAL CIRCULATIONS}

Prominent meridional circulations are induced by the convection within the
unstable core, accompanied by much weaker circulations in the radiative
surroundings.  Within the convective core, such mean meridional flows arise
from a combination of buoyancy forces, Reynolds stresses, latitudinal
pressure gradients, and of Coriolis forces acting on the mean zonal flow
(differential rotation).  In the radiative exterior beyond the zone of
overshooting, slow circulations can be driven both by viscous stresses that
couple that region to the flows in the convective interior, and by
latitudinal temperature gradients arising from the nonuniform heating from
below.  Our simulations reveal that the resulting meridional circulations
are complex in structure, possess multiple cells in both radius and
latitude, and exhibit noticeable variations in time.  Such multi-cellular
patterns are in sharp contrast to the single monolithic cells often assumed
in mean-field models of differential rotation (e.g., Rekowski \& R\"udiger
1998; Durney 2000).  On the other hand, our models include viscous stresses
that may be largely absent in real stars, and thus our estimates of the meridional
circulations in the immediate surroundings of the convective core are
likely to be somewhat unrealistic.

\subsection {Patterns Within Convective Core}

The time-averaged meridional circulations achieved in three of our
simulations are shown in Figure $14$ as streamfunctions of mass flux
(cf. equation [7] of Miesch et al. 2000).  Within the convective core (left
panels), these flows have amplitudes of order $5$ m s$^{-1}$; they are thus
appreciably smaller than the instantaneous radial velocities in the same simulations
(cf. Table 2).  Indeed, the kinetic energy of these axisymmetric meridional
flows (MCKE) is in all cases at least an order of magnitude smaller than
that of the non-axisymmetric convection (CKE) and of the differential
rotation (DRKE).  More rapid flows, and more intricate cellular patterns,
are realized in the less viscous cases C and C4 than in case B.

The patterns of meridional flow are quite intricate, with multi-celled
structures achieved in radius and latitude.  These structures exhibit some
alignment with the rotation axis, particularly in the more rapidly rotating
case C4. As the convection itself evolves (Fig. $8$), so do the mean
meridional circulations associated with it.  Thus Figure $14$ represents
typical examples of these time-dependent circulations in a particular
epoch.  These circulations occupy most of the volume delineated by $r_c$,
and weaken in the region of overshooting.  Thus their extent varies with
the stiffness of the exterior stratification.

\subsection {Weak External Circulations}

The meridional circulations achieved in the radiative envelope within our
models are much weaker than those in the core.  As shown in the right
panels of Figure $14$, these flows are of large scale and possess
amplitudes that in cases B and C are about $0.06$ m s$^{-1}$.  The more
rapidly rotating case C4 exhibits faster exterior circulations.  Such flows
evolve more slowly than do the meridional circulations within the core.
They are driven by a combination of viscous stresses communicated across
these domains (with a characteristic diffusion timescale across the
radiative envelope of about $1.8$y for case C) and by weak latitudinal
thermal gradients resulting from the energy deposited by the
latitudinally-varying $\hat{F_e}$ (Fig. $11$).

These mean flows within the radiative envelopes in our models are
considerably greater in amplitude than estimates of Eddington-Sweet
circulations.  The latter, which arise from baroclinic effects due to the
flattened shape of a rotating star, would involve flows of order $10^{-4}$
m s$^{-1}$ in our A-type stars rotating at $\Omega_o$ (see review in
Pinsonneault 1997).  Our modeling ignores centrifugal acceleration and thus
such distortions, and further imposes a spherical outer boundary upon which
a uniform $d S/dr$ is specified.  Therefore our model does not accurately
address the very slow circulations of the Eddington-Sweet variety.


\section {INTERPRETING THE ROTATION PROFILES}

Our simplified model of rotating core convection coupled by penetration to
a radiative envelope yields a nonlinear dynamical system with complex
feedbacks.  The rich variety of responses and marked time variability that
is exhibited cannot now be predicted or explained from first principles.
However, the numerical simulations have the great advantage that they
provide full snapshots that can be assessed for the dynamical balances
that are achieved.  This can yield insights into the operation of these
complicated dynamical systems.  Such an approach is particularly useful in
identifying the processes involved in continuously redistributing the
angular momentum to establish the differential rotation. Because we have
adopted stress-free boundary conditions at the top and bottom boundaries of
our simulations, no net torque is applied to these rotating spheres of
convection.  Thus total angular momentum within our simulations is
conserved, and we now examine the manner in which it is redistributed.

\subsection {Redistributing the Angular Momentum}

We may assess the transport mechanisms of
angular momentum in our simulations, which must combine to give rise to the
observed differential rotation, in the manner of Elliott et al. (2000) and
Brun \& Toomre (2002).  We consider the mean radial (${\cal F}_r$) and
latitudinal (${\cal F}_{\theta}$) angular momentum fluxes, and write the
$\phi$-component of the momentum equation, expressed in conservative form
and averaged in time and longitude, as
\begin{equation}
\frac{1}{r^2} \frac{\p(r^2 {\cal F}_r)}{\p r}+\frac{1}{r \sin\theta}
\frac{\p(\sin \theta {\cal F}_{\theta})}{\p
\theta}=0,
\end{equation}
involving the mean  radial angular momentum flux
\begin{equation}
{\cal F}_r=\hat{\rho}r\sin\theta[-\nu r\frac{\p}{\p
r}\left(\frac{\hat{v}_{\phi}}{r}\right)+\widehat{v_{r}^{'}
v_{\phi}^{'}}+\hat{v}_r(\hat{v}_{\phi}+\Omega r\sin\theta)] \end{equation}
and the mean latitudinal angular momentum flux
\begin{equation}
{\cal F}_{\theta}=\hat{\rho}r\sin\theta[-\nu
\frac{\sin\theta}{r}\frac{\p}{\p
\theta}\left(\frac{\hat{v}_{\phi}}{\sin\theta}\right)+\widehat
{v_{\theta}^{'} v_{\phi}^{'}}+\hat{v}_{\theta}(\hat{v}_{\phi}+\Omega
r\sin\theta)].
\end{equation}
In the above expressions for both fluxes, the first terms in each
bracket are related to the  angular  momentum  flux due  to  viscous
transport (which we denote as  ${\cal  F}_{r,V}$ and ${\cal
F}_{\theta,V}$), the second term to the transport due  to  Reynolds
stresses (${\cal F}_{r,R}$ and ${\cal F}_{\theta,R}$) and the  third
term to the transport by the meridional circulation (${\cal F}_{r,M}$
and  ${\cal F}_{\theta,M}$).  The Reynolds stresses in the above
expressions arise through correlations of the components of velocity.    

In Figure $15$, we show the components of ${\cal F}_r$ and ${\cal
F}_{\theta}$ for cases $D$ and $E$ with a strong rotational influence,
integrated along co-latitude and radius respectively to deduce the net
fluxes through shells at various radii and through cones at various
latitudes, such that
\begin{equation}
I_{{\cal F}_r}(r)=\int_0^{\pi} {\cal F}_r(r,\theta) \, r^2 \sin\theta
\, d\theta \; \mbox{ , } \; I_{{\cal
F}_{\theta}}(\theta)=\int_{r_{bot}}^{r_{top}} {\cal
F}_{\theta}(r,\theta) \, r \sin\theta \, dr \, ,
\end{equation}
We have identified  the contributions from viscous (V), Reynolds
stresses (R) and meridional circulation (M) terms. This representation
is helpful in considering the sense and amplitude of the transport of
angular  momentum within the convective shells by each component of
${\cal F}_r$ and ${\cal F}_{\theta}$.

Turning first to the radial fluxes of angular momentum (Fig. $15a$, $b$),
we see that viscous forces act to transport angular momentum radially
inward in both cases.  In this they are opposed by a combination of the
Reynolds stress and meridional circulation fluxes, to yield a total radial
angular momentum flux that is nearly zero, as noted in Fig. $15$ by the solid
line.  While the systems here are highly variable in time, by allowing the
system to evolve for extended periods of time (typically thousands of days)
and performing long time averages, we appear to be sensing the equilibrated
states reasonably well. In going from the more laminar flows of case $D$ to
the complex ones of case $E$, we see that the viscous flux has dropped and
that the Reynolds stresses and meridional circulations have changed
accordingly to maintain equilibrium.  This includes the development of an
appreciable integrated ${\cal F}_{r,M}$ that is directed radially inward,
near the boundary of the radiative and convective zones; this is presumably
a result of the stronger circulations and shear generated near that
boundary by the more complex flows of case $E$.

Examining the latitudinal transport of angular momentum (Fig. 15$c$, $d$),
we note that the effect of the Reynolds stresses in both cases is primarily
to speed up the equator, since ${\cal F}_{\theta,R}$ is positive in the
northern hemisphere and negative in the southern.  It is opposed by the
meridional circulation and viscous fluxes, which act to speed up the
poles.  The manner in which each of the different components of ${\cal
F}_{\theta}$ acts does not appear to vary appreciably in going from case $D$
to $E$.  As the level of complexity is increased, however, we see a
decrease in the magnitude of all components of ${\cal F}_{\theta}$.

We conclude that for these cases involving a strong rotational constraint,
the Reynolds stresses act latitudinally to speed up the equator, and
radially to slow down the convective core.  These latidudinal fluxes are
opposed by the meridional circulations and viscous effects, whereas the
radial fluxes are aided somewhat by the meridional circulations.  Examining
these fluxes gives us a picture of how the angular momentum is continuously
transported, and which are the significant processes serving to speed up or
slow down certain regions.  However, it is difficult to infer from the
fluxes alone what will be the detailed $\Omega$ profile that results.

\subsection {Baroclinicity and Thermal Winds}

Rotating convection involves both radial and latitudinal heat transport,
with the likelihood that latitudinal gradients in temperature and entropy
may result within the convective core.  This implies that surfaces of mean
pressure and density will not coincide, thereby yielding baroclinic terms
in the vorticity equations (Pedlosky 1987; Zahn 1992).  Under sufficiently
strong rotational constraints, a `thermal wind balance' might be achieved
in which departures of the angular velocity from being constant on
cylinders (aligned with the rotation axis) are controlled by those
baroclinic terms.  Indeed, some mean-field approaches have invoked such a
balance to obtain differential rotation profiles with bearing on the solar
convection zone (e.g., Kichatinov \& R\"udiger 1995). As discussed in Brun
\& Toomre (2002), such a balance effectively implies that
\begin{equation}
\frac{\partial\hat{v}_{\phi}}{\partial z} =
\frac{1}{2
\Omega_o\hat{\rho}c_{P}}\del\,\hat{S}\,\dotp\del\,\hat{p}\Bigr|_{\phi}
= \frac{g}{2 \Omega_o r\,c_{P}} \frac{\partial\hat{S}}{\partial \theta}
\mbox{ ,}\label{tw2}\end{equation}
where $z$ is parallel to the rotation axis.  Thus latitudinal entropy
gradients could serve to break the Taylor-Proudman constraint, which would
otherwise require the rotation to be constant on cylinders with
$\partial\hat{v}_{\phi}/\partial z$ equal to zero.  That constraint may
also be broken by Reynolds and viscous stresses.  We next show that these
terms are at least as important as the baroclinic terms in establishing the
differential rotation in the bulk of the convective core.

Figure $16$ assesses for case E the extent to which latitudinal entropy
gradients serve to drive the temporal mean zonal flows $\hat{v}_{\phi}$
seen as the differential rotation.  Figure $16b$ displays the gradient
$\partial\hat{v}_{\phi}/\partial z$ of those zonal flows, and Figure $16c$
the corresponding latitudinal entropy term on the right hand side of
equation (\ref{tw2}), which in an exact thermal wind balance would be identical.
Figure $16d$ shows the difference between this baroclinic term and the
actual $\partial\hat{v}_{\phi}/\partial z$, thus providing a measure of
departures from such a balance.  Within the bulk of the convective core,
Reynolds and viscous stresses lead to zonal flows that are not so balanced.
At the interface between the core and the radiative exterior, however,
baroclinicity accounts for most of the differential rotation.

Thus examining the nature of this thermal wind balance, combined with the
assessment of fluxes of angular momentum, reveals that Reynolds and viscous
stresses have a major role in establishing the differential rotation in the
bulk of the core, with latitudinal thermal gradients influencing the
profile near the interface between the convective region and the radiative
envelope.  

\section {SUMMARY AND PERSPECTIVES}

Our highly simplified treatment of core convection within A-type stars has
begun to capture some of the complex dynamics that must be occurring when
turbulent convection and rotation meet in a full spherical domain.  The
most severe simplifications in our models are that we are able to simulate
only the central portions of these stars, that compositional gradients that
must be present are ignored, and that the levels of turbulence studied here
are only modest compared to what may be realized in real stars.  The
dynamics may well be influenced by each in turn, and in ways that are
difficult to predict at this stage.  Though our work is thus quite
preliminary, it has revealed several dynamical properties that may well
turn out to be robust.

Firstly, these convective cores rotate differentially.  Prominent angular
velocity contrasts are established in all cases studied here, with a
central column of slow rotation realized in stars that are strongly
influenced by rotation (with $R_{oc}$ small).  Within the bulk of the
convection zone, this differential rotation is driven primarily by Reynolds
and viscous stresses, whereas near the interface between the core and the
radiative envelope, baroclinicity plays an important role.

Secondly, convective motions penetrate into the radiative envelope,
yielding a prolate shape to the region of nearly adiabatic stratification.
The surrounding region of overshooting, in which convective motions persist
but do not appreciably modify the prevailing entropy gradient, is found to
have an outer boundary that is largely independent of latitude, yielding a
spherical shape to the domain experiencing convective stirring.

Thirdly, small-amplitude circulations and gravity waves are excited within
the radiative envelope.  The latter appear in some simulations as a
rippling patchwork of waves, and in our more rapidly rotating case
C4 as a striking global-scale resonance that once established, appears to persist.

The simplifications we have made in our modeling are likely to impact the
penetrative properties to some degree.  In particular, the $\mu$-gradients
of chemical composition present in real stars should diminish the extent of
penetrative and overshooting motions realized there (Zahn 1991).  They also
probably reduce the vigor of circulations within the radiative envelope
(Lebreton \& Maeder 1987).  More turbulent flows are likely to yield
smaller spatial filling factors for the convective plumes, and hence
probably also lessen the extent of overshooting (Brummell et al. 2002).
Further, real stars possess stiffer entropy stratifications in their
radiative envelopes than those we have been able to consider.  This too
would lead to less overshooting.  Thus our estimates here provide upper
limits for the extent of penetration and overshooting that may be achieved
in real stars.

The strong regions of shear realized in all our simulations may contribute
to the building of orderly magnetic fields by dynamo action.  Poloidal
fields can be converted into toroidal ones by such differential rotation
(the $\Omega$-effect), with regeneration of the poloidal fields perhaps
accomplished by the helical turbulence (the $\alpha$-effect; cf. Mestel
1999).  Acting together, these processes may build well-organized fields
that could, if subject to magnetic buoyancy instabilities, rise to the
surfaces of these stars where they might be observed (MacGregor \&
Cassinelli 2003).  We are in the process of beginning to explore the
generation of magnetic fields within these cores (cf. Browning, Brun, \& Toomre
2003).

The penetrative convection and overshooting in our simulation may likewise
have some observable consequences.  Although we have not investigated in
any detail the bringing of fresh fuel into the convective core by these
motions, some replenishment of the hydrogen stockpile available for nuclear
reactions must occur, and thereby lead to a prolonged lifetime on the
main-sequence.  

We thank Douglas Gough and Jean-Paul Zahn for very helpful discussions.
J.T. thanks the Observatory of Paris, Meudon and the CEA-Saclay for their
hospitality during the interpretation of these simulations.  This work was
partly supported by NASA through SEC Theory Program grant NAG5-11879, and
through the Graduate Student Researchers Program (NGT5-50416).  Various
phases of the simulations here were carried out with NSF PACI support of
the San Diego Supercomputing Center (SDSC), the National Center for
Supercomputing Applications (NCSA), and the Pittsburgh Supercomputing
Center (PSC).  Much of the analyses of the extensive data sets were
conducted in the Laboratory for Computational Dynamics (LCD) within JILA.

\begin{deluxetable}{llllllll}
\tablecolumns{8}
\tablenum{1}
\tablecaption{Summary of simulation parameters}
\tablehead{
\colhead{} & \colhead{A} & \colhead{B} & \colhead{C} & \colhead{D} & \colhead{E}
& \colhead{C4} & \colhead{F}
}
\startdata
$\Omega_o/\Omega_{\sun}$ & 1 & 1 & 1 & 1 & 1 & 4 & 0.1\\
max d$\sb$/dr & 1.5$\times 10^{-6}$ & 1.5$\times 10^{-5}$ & 1.5$\times 10^{-5}$
& 1.5$\times 10^{-4}$ & 1.5$\times 10^{-4}$ & 1.5$\times 10^{-5}$ &
1.5$\times 10^{-4}$\\
$\nu$ & $4.35 \times 10^{11}$ & $2.49 \times 10^{12}$ &
$4.35 \times 10^{11}$ & $2.49 \times 10^{12}$ & $4.35 \times 10^{11}$ &
$2.49 \times 10^{11}$ & $4.35 \times 10^{11}$ \\
$\kappa$ & $1.74 \times 10^{12}$ & $9.96 \times 10^{12}$ &
$1.74 \times 10^{12}$ & $9.96 \times 10^{12}$ & $1.74 \times 10^{12}$ &
$9.96 \times 10^{11}$ & $1.74 \times 10^{12}$\\
$R_a$ & 1.7$\times 10^5$ & 2.0$\times 10^4$ & 1.9$\times 10^5$ & 1.0$\times
10^4$ & 3.3$\times 10^5$ & 2.2$\times 10^6$ & 3.3$\times 10^5$\\
$T_a$ & 1.2$\times 10^7$ & 3.6$\times 10^5$ & 1.2$\times 10^7$
& 3.6$\times 10^5$ & 1.2$\times 10^7$ & 5.9$\times 10^8$ & 1.2$\times
10^5$\\
$R_{oc}$ & 0.24 & 0.47 & 0.25 & 0.33 & 0.33 & 0.12 & 3.3\\
$R_e$ & 1.7$\times 10^2$ & 1.2$\times 10^1$ & 1.3$\times 10^2$ &
1.1$\times 10^1$ & 1.4$\times 10^2$ & 2.0$\times 10^2$ & 1.7$\times 10^2$
\enddata

\tablecomments{All simulations have an inner radius of $3.0 \times 10^{9}$
  cm and an outer radius of $4.0 \times 10^{10}$ cm, with $L=1.7 \times
  10^{10}$ cm the approximate radial extent of the convective core. The
  Prandtl number $P_r=\nu/\kappa = 0.25$ in all cases.  Here evaluated at
  mid-layer depth are the Rayleigh number $R_a=(-\p \rb/\p \sb)\Delta S g
  L^3/\rho \nu \kappa$ (with $\Delta S$ the entropy contrast across the
  core), the Taylor number $T_a=4 \Omega^2 L^4/\nu^2$, the convective
  Rossby number $R_{oc}=\sqrt{R_a/T_a P_r}$, and the rms Reynolds number
  $\tilde{R}_e=\vvr L/\nu$, where $\vvr$ is a representative rms convective
  velocity. A Reynolds number based on the peak velocity at mid depth would
  be about fourfold larger. The eddy viscosity $\nu$ and eddy conductivity
  $\kappa$ at the middle of the core are quoted in cm$^2\,$s$^{-1}$.  }
\end{deluxetable}

\begin{deluxetable}{lllllll}
\tablecolumns{7}
\tablenum{2}
\tablecaption{Velocities, energies, differential rotation, and overshooting}
\tablehead{
\colhead{} & \colhead{A} & \colhead{B} & \colhead{C} & \colhead{D} & \colhead{E}
& \colhead{C4}
}
\startdata
$\vrr$ & 36 & 15 & 28 & 14 & 30 & 26 \\
$\vtr$ & 27 & 24 & 26 & 22 & 26 & 21 \\
KE & 1.0$\times 10^8$ & 3.3$\times 10^7$ & 7.6$\times 10^7$ & 2.2$\times 10^7$ &
	7.4$\times 10^7$ & 1.2$\times 10^8$\\
DRKE & 5.6$\times 10^7$ & 3.5$\times 10^6$ & 4.6$\times 10^7$ & 5.6$\times 10^6$ &
	4.3$\times 10^7$ & 1.0$\times 10^8$\\
CKE & 4.2$\times 10^7$ & 2.9$\times 10^7$ & 2.9$\times 10^7$ & 1.6$\times 10^7$ &
	3.0$\times 10^7$ & 1.3$\times 10^7$\\
MCKE & 2.5$\times 10^6$ & 5.4$\times 10^4$ & 8.6$\times 10^5$ & 2.4$\times 10^4$ &
	1.1$\times 10^6$ & 3.7$\times 10^5$\\
$\Delta\Omega/\Omega_o$ &
     65\% & 26\% & 56\% & 28\% & 62\% & 36\% \\
$\bar{d_o}$ &
     3.41$\times 10^9$ & 3.35$\times 10^9$ & 2.20$\times 10^9$
   & 3.25$\times 10^9$ & 1.75$\times 10^9$ & 2.57$\times 10^9$ \\
$r_c^e$ &
     1.93$\times 10^{10}$ & 1.82$\times 10^{10}$ & 1.80$\times 10^{10}$
   & 1.89$\times 10^{10}$ & 1.76$\times 10^{10}$ & 1.89$\times 10^{10}$ \\
$r_c^p$ &
     2.16$\times 10^{10}$ & 1.97$\times 10^{10}$ & 2.05$\times 10^{10}$
   & 2.03$\times 10^{10}$ & 2.00$\times 10^{10}$ & 2.05$\times 10^{10}$ \\
$\bar{r_o}$ &
     2.34$\times 10^{10}$ & 2.24$\times 10^{10}$ & 2.19$\times 10^{10}$
   & 2.10$\times 10^{10}$ & 2.07$\times 10^{10}$ & 2.28$\times 10^{10}$ \\
\enddata
\tablecomments{Components of rms velocity components $\vrr$ and $\vtr$ at
mid-core depth ($r=0.10 R$) are expressed in m s$^{-1}$.  The
temporal averages over the full domain of the kinetic energy density (KE),
and the energies associated with the differential rotation
(DRKE), convection (CKE), and meridional circulation (MCKE), are all expressed
in erg cm$^{-3}$.  The relative angular velocity contrasts
$\Delta\Omega/\Omega_o$ between $0$\degr and $60$\degr \ latitude are stated.  The
geometry of the prolate convective core and of the surrounding region of
overshooting is measured by the radius of the convectively
unstable region at the equator ($r_c^e$) and at the poles ($r_c^p$), the
outer boundary of the region of overshooting ($\bar{r_o}$), and the spherical
average of the spatial extent of the overshoot ($\bar{d_o}$), all in cm.
}
\end{deluxetable}

\clearpage

\begin{figure}[hpt]
  \center
  \epsscale{1.0}
  \plotone{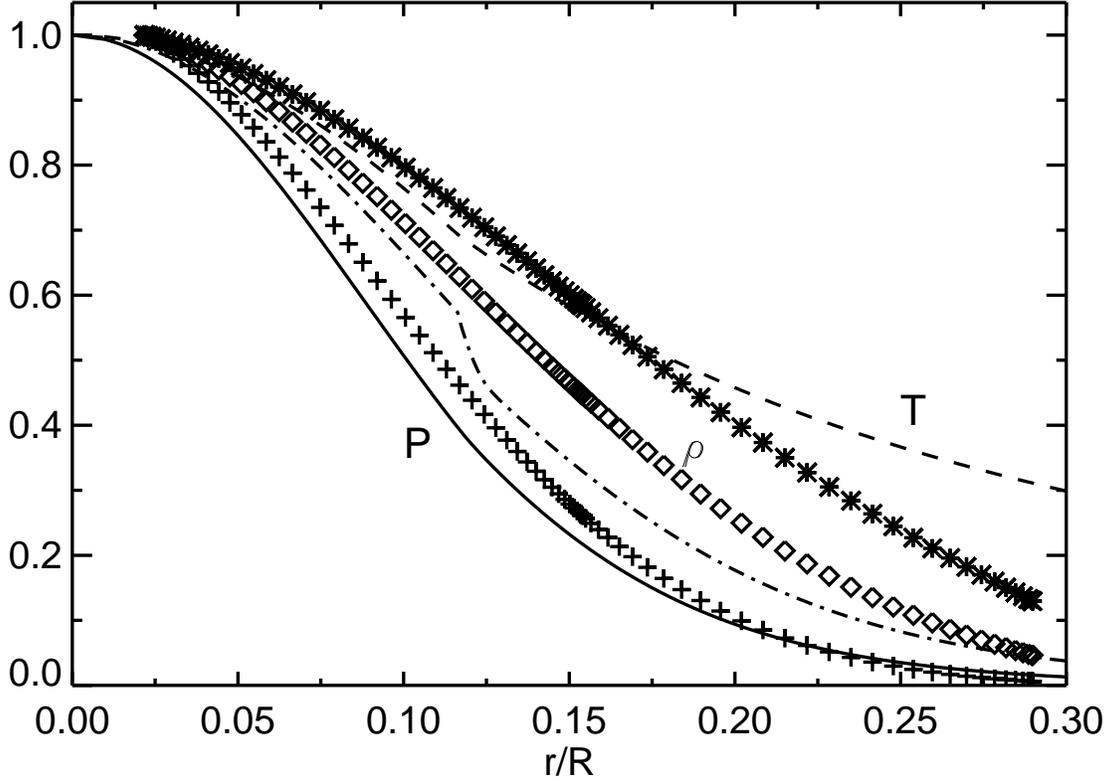}
  \caption{Mean radial stratification in the central regions of a 2 solar
  mass A-type star, within both a 1--D stellar structure model and case C.
  The mean pressure $\bar{P}$, temperature $\tb$, and density $\rb$ are
  plotted with symbols at their radial mesh locations for case C, and as
  continuous curves for the 1--D model.  All variables are normalized to
  their maximum values, which for case C at the inner boundary of the
  computational domain ($r=0.022 R$) are $\bar{\rho}_{o} = 48$ g cm$^{-3}$,
  $\bar{T}_{o}=2.0 \times 10^{7}$ K, and $\bar{P}_{o} = 1.4 \times 10^{17}$ dynes
  cm$^{-2}$}.
\end{figure}

\begin{figure}[hpt]
  \center
  \epsscale{1.0}
  \plottwo{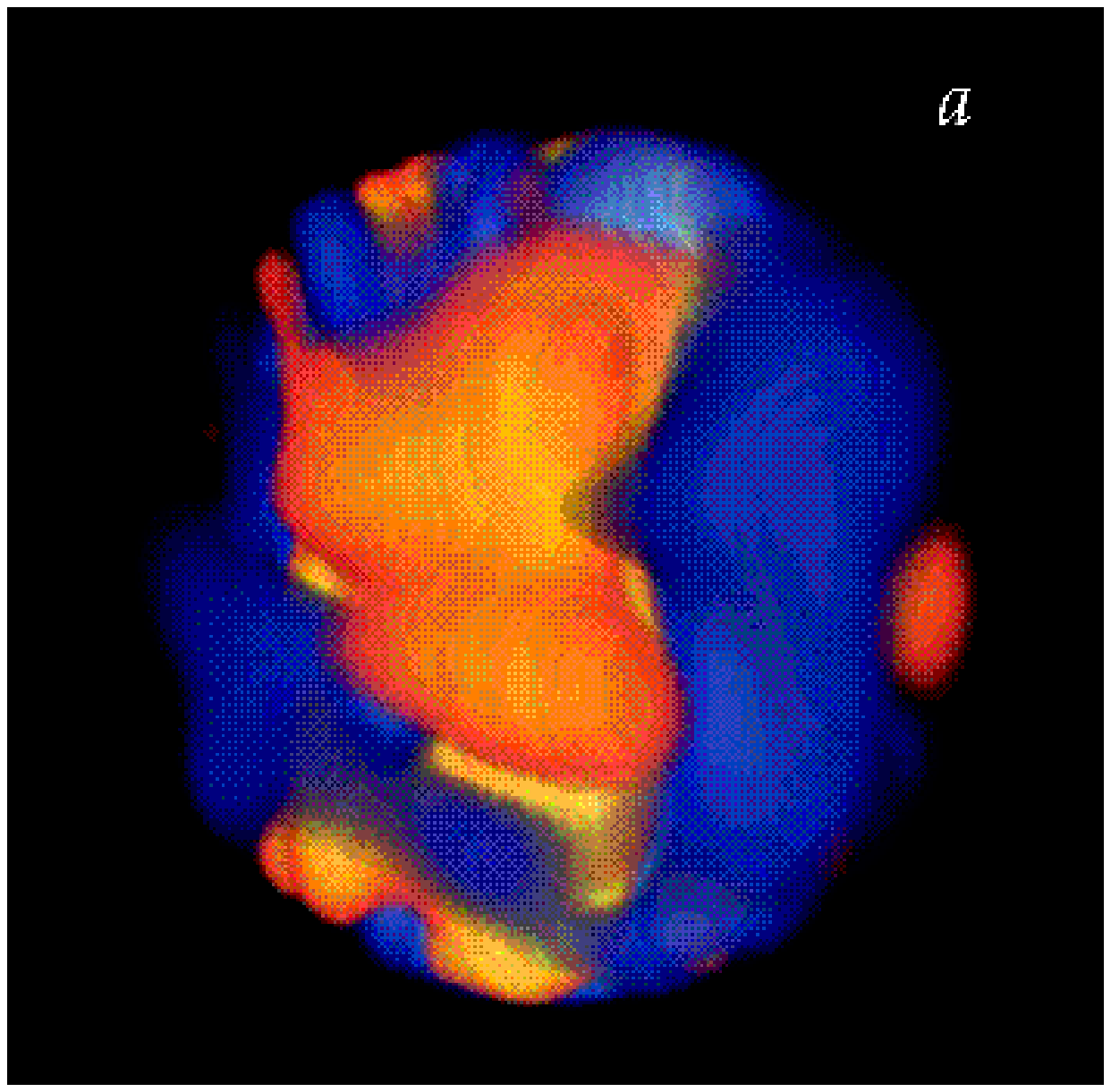}{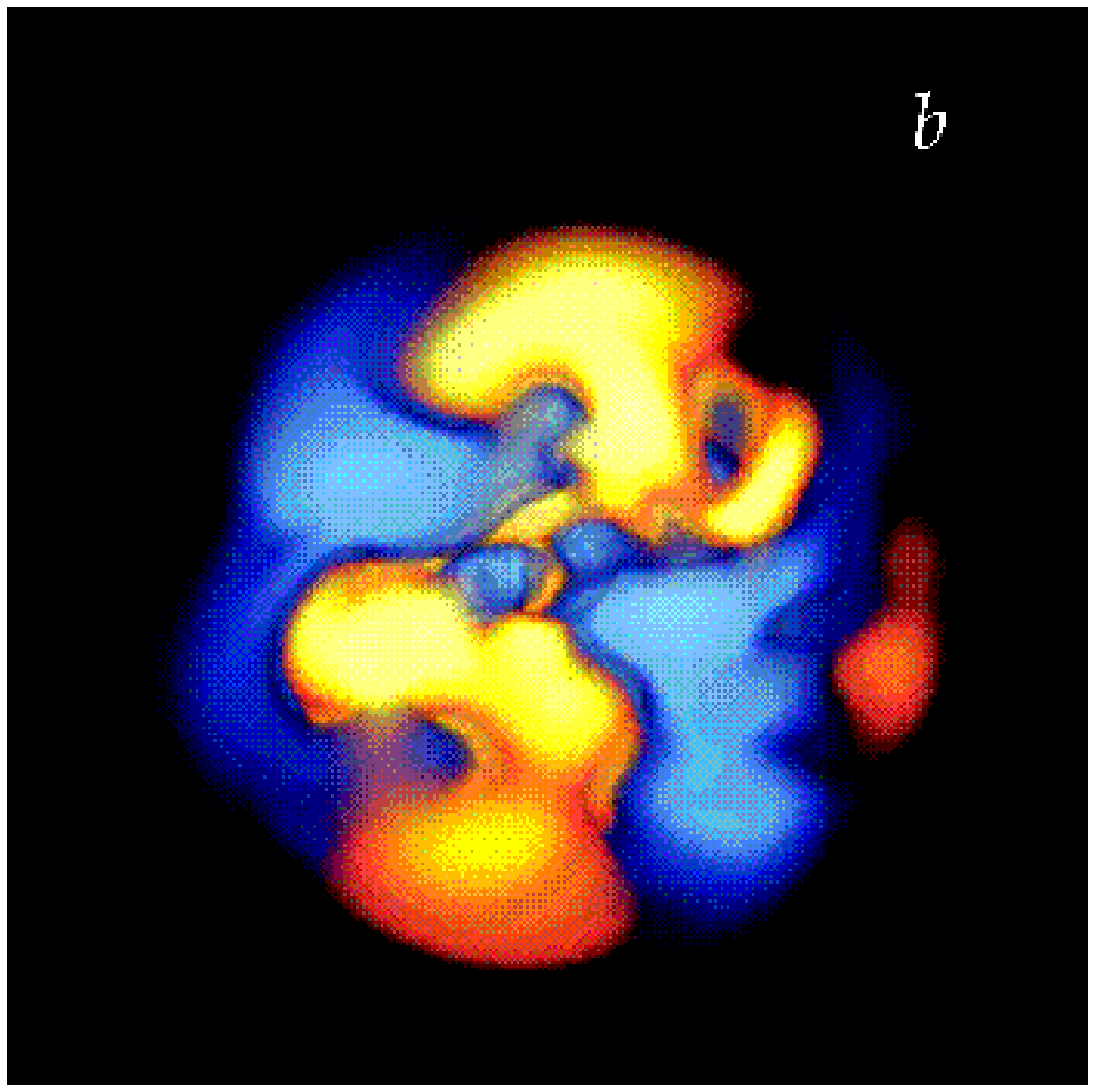}
  \caption{Radial velocity at one instant in time for case E, using volume
  rendering of flow structure ($a$) near the outer boundary of the prolate
  convective core, with the rotation axis oriented vertically, and ($b$)
  near the equatorial plane viewed from above. Upflows (radially outward) are rendered
  as yellow/red tones; downflows are bluish.}
\end{figure}

\begin{figure}[hpt]
  \center
  \epsscale{1.0}
  \plotone{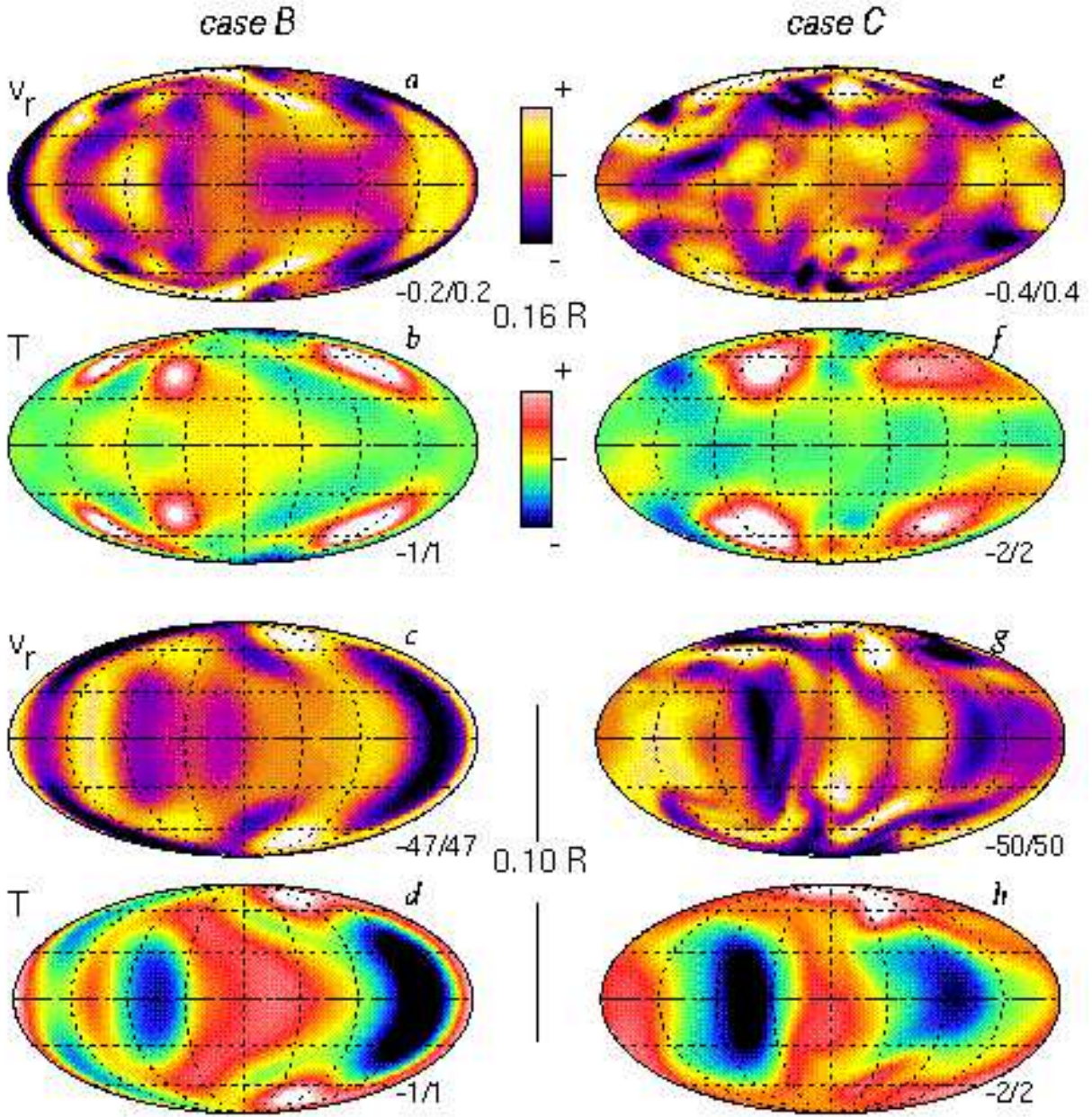}
  \caption{Global mappings of radial velocities ($v_r$) and associated
  temperature fluctuations ($T$) at one instant in time for the laminar
  case B (left column) and the more complex case C (right column), shown in
  Mollweide projection with the dashed horizontal line designating the
  equator.  Two depths are sampled, with the lower panels exhibiting $v_r$
  and $T$ within the convective core (at $r=0.10R$) and the upper panels
  within the region of overshooting (at $r=0.16R$).  One color rendering is
  used for $v_r$, and another for $T$, with the maximum and minimum
  amplitudes for each field (in m s$^{-1}$ and K) indicated next to each panel.}
\end{figure}

\begin{figure}[hpt]
  \center
  \epsscale{1.0}
  \plotone{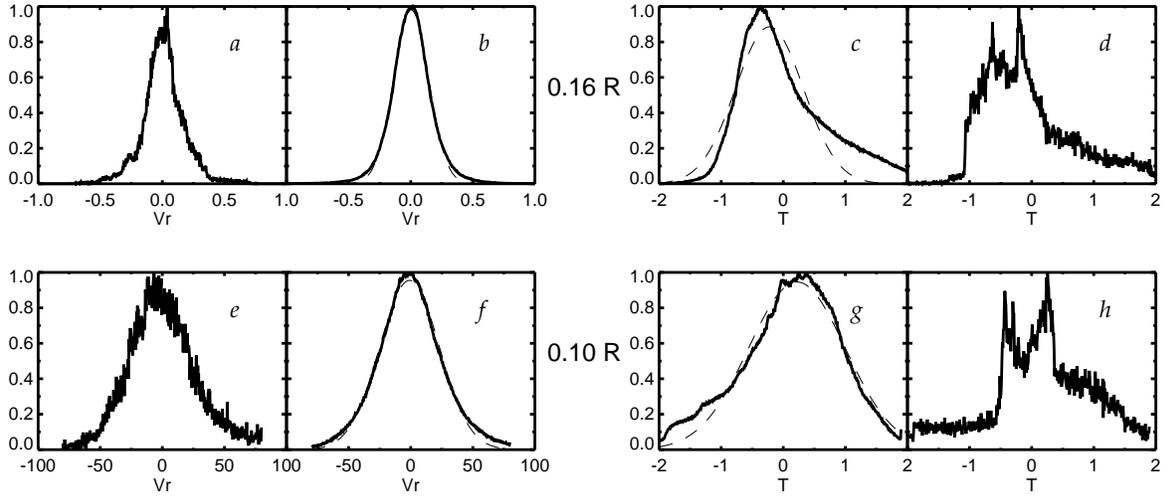}
  \caption{Probability distribution functions in case C for radial
  velocities (left panels)
  and temperature fluctuations (right) within the convective core ($e-h$,
  at $r=0.10 R$) and the region of overshooting ($a-d$, at $r=0.16 R$).
  Shown in ($a,d,e,h$) are instantaneous pdfs for the two fields; ($b,c,f,g$)
  depict temporal averages over about 180 days together with the best-fit Gaussian of zero mean
  (shown dashed).  The distribution of radial velocities very closely
  approximates a Gaussian, whereas the fluctuations of temperature are
  slightly asymmetric about their most probable value.}
\end{figure}


\begin{figure}[hpt]
  \center
  \epsscale{1.0}
  \plotone{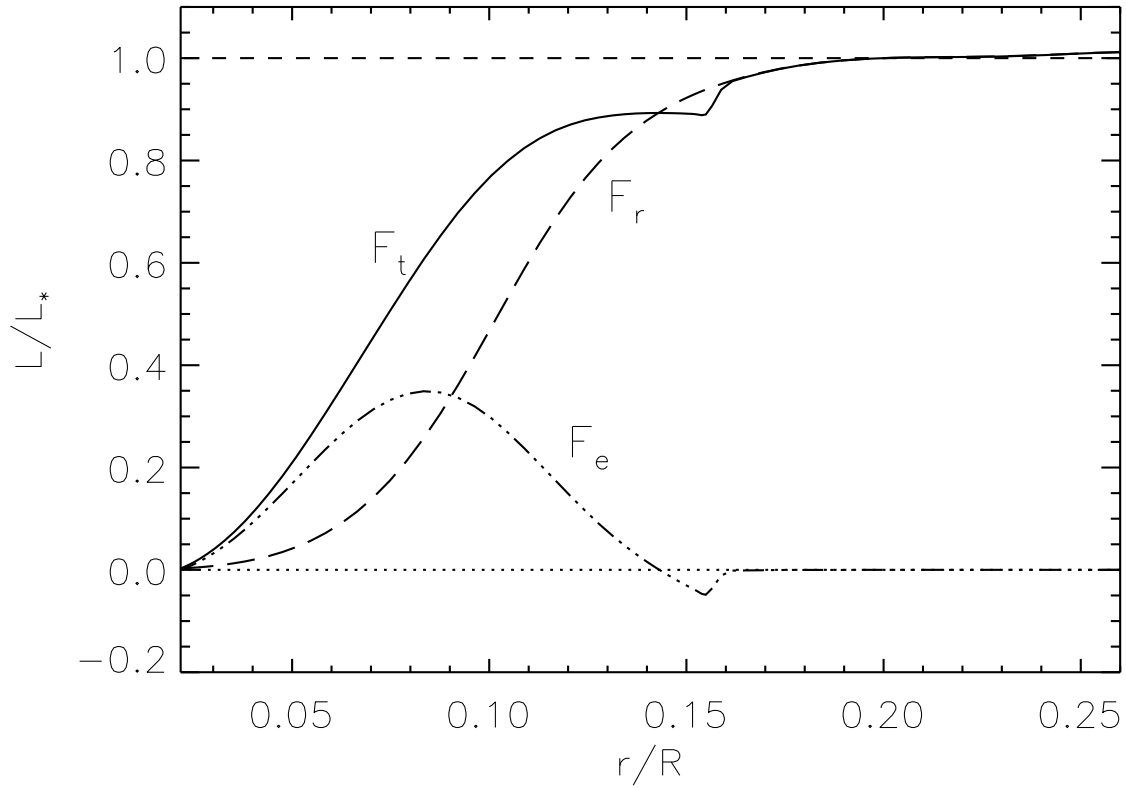}
  \caption{Time-averaged radial transport of energy within case C as a
function of proportional radius, formed over an interval of about 180
days. Shown are the enthalpy flux $F_{e}$ and the radiative flux $F_{r}$,
together with their sum, the total flux $F_{t}$; all quantities have been
expressed as luminosities.  The convective core extends here to about
$r=0.14R$, with the positive $F_{e}$ there serving to carry as much as 57\%
of the total flux.  The further region of overshooting involves a small
negative (inward directed) enthalpy flux.}
\end{figure}

\begin{figure}[hpt]
  \center
  \epsscale{0.5}
  \plotone{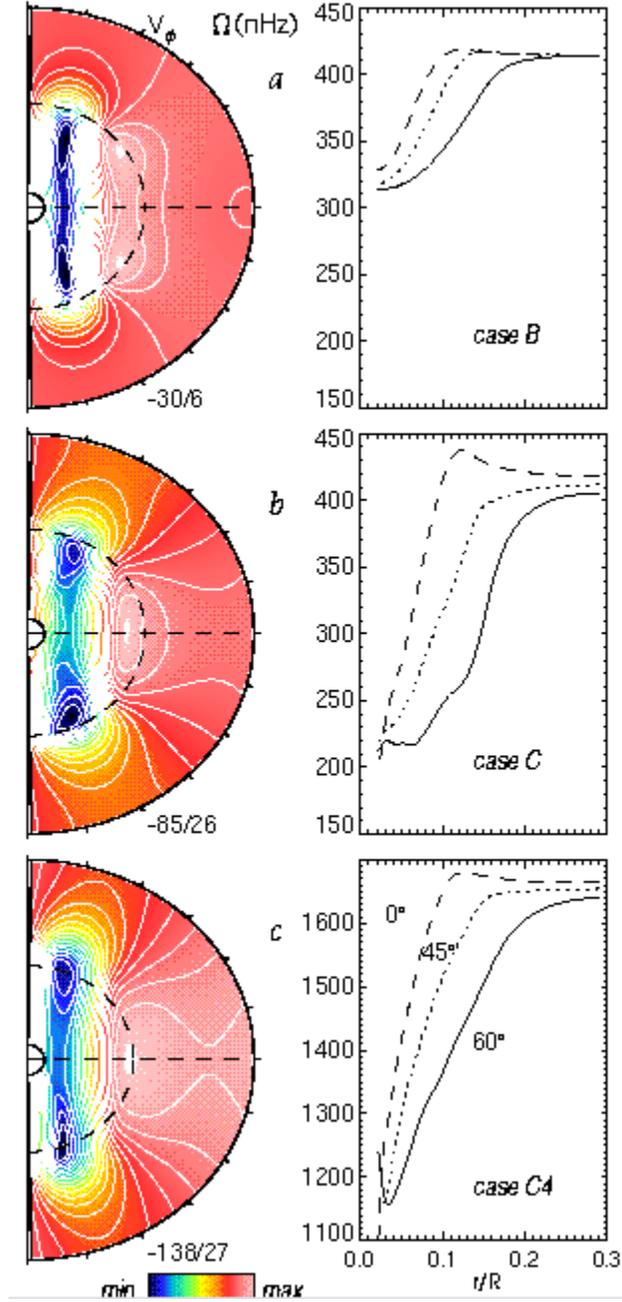}
  \caption{Strong differential rotation established by the convection in
  the three cases B, C, and C4.  Shown are the azimuthal velocity $\vph$ as
  contour plots in radius and latitude (left panels, with dashed line denoting
  equator), and the angular velocity $\hat{\Omega}$ with proportional
  radius for three latitudinal cuts (right panels); both $\vph$ and
  $\hat{\Omega}$ are averaged in time (all over intervals of roughly 110
  days) and longitude.  Maxima and minima of
  $\vph$ (in m s$^{-1}$) are indicated.  Cases B and C (upper, middle
  panels) are rotating at the solar rate (414 nHz), and case C4 (lower
  panels) rotates fourfold faster.  Greater angular velocity contrasts are
  achieved in the more complex flows of case C, C4 ($b$,$c$) than in the
  laminar case B ($a$), in all cases involving a central column of
  slowness.}
\end{figure}

\begin{figure}[hpt]
   \center
   \epsscale{0.5}
   \plotone{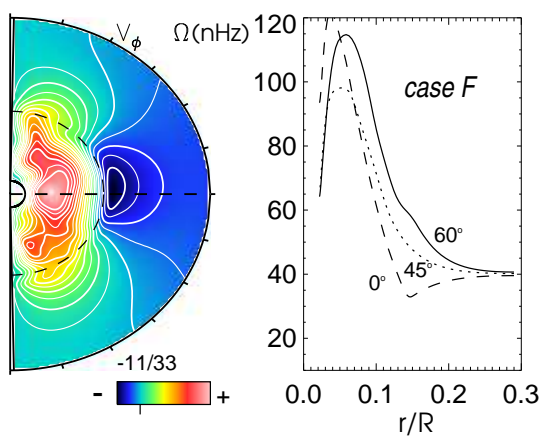}
   \vspace{-3.0in}
   \caption{Fast column of central rotation realized in case F rotating at
  one-tenth the solar rate.  Shown is the azimuthal velocity $\vph$ as a
  contour plot in radius and latitude (left panel, with dashed line denoting
  equator), and the angular velocity $\hat{\Omega}$ with proportional
  radius for three latitudinal cuts (right panel); both $\vph$ and
  $\hat{\Omega}$ are averaged in time (over 180 days) and longitude.
   Maxima and minima of $\vph$, in m s$^{-1}$, are indicated.}  
\end{figure}

\begin{figure}[hpt]
  \center
  \epsscale{1.0}
  \plotone{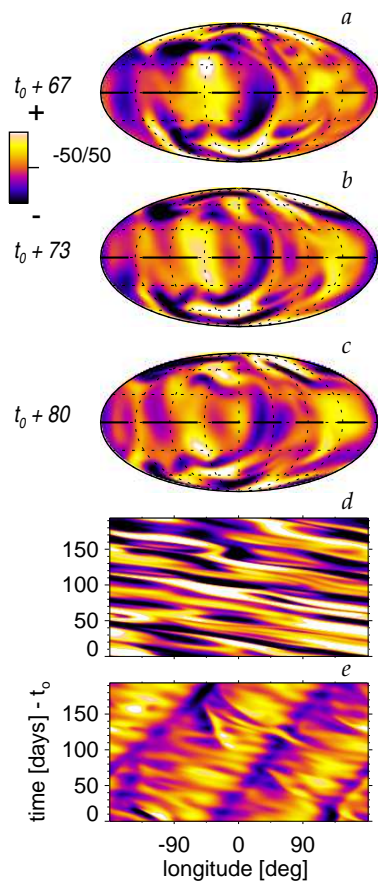}
  \caption{Evolution and propagation of the convective flows sampled for
  case C.  ($a-c$) Global mappings of the radial velocity $v_r$ within the
  convective core (at $r=0.10R$) for three successive times each separated
  by 7 days, exhibiting shearing and cleaving of the convective
  structures.  The propagation of those convective patterns is most evident
  in the time-longitude maps of $v_r$ extracted at ($d$) a latitude of 65\degr
  \  and ($e$) at the equator.  The same color rendering is used in
  all panels.}
\end{figure}

\begin{figure}[hpt]
  \center
  \epsscale{1.0}
  \plotone{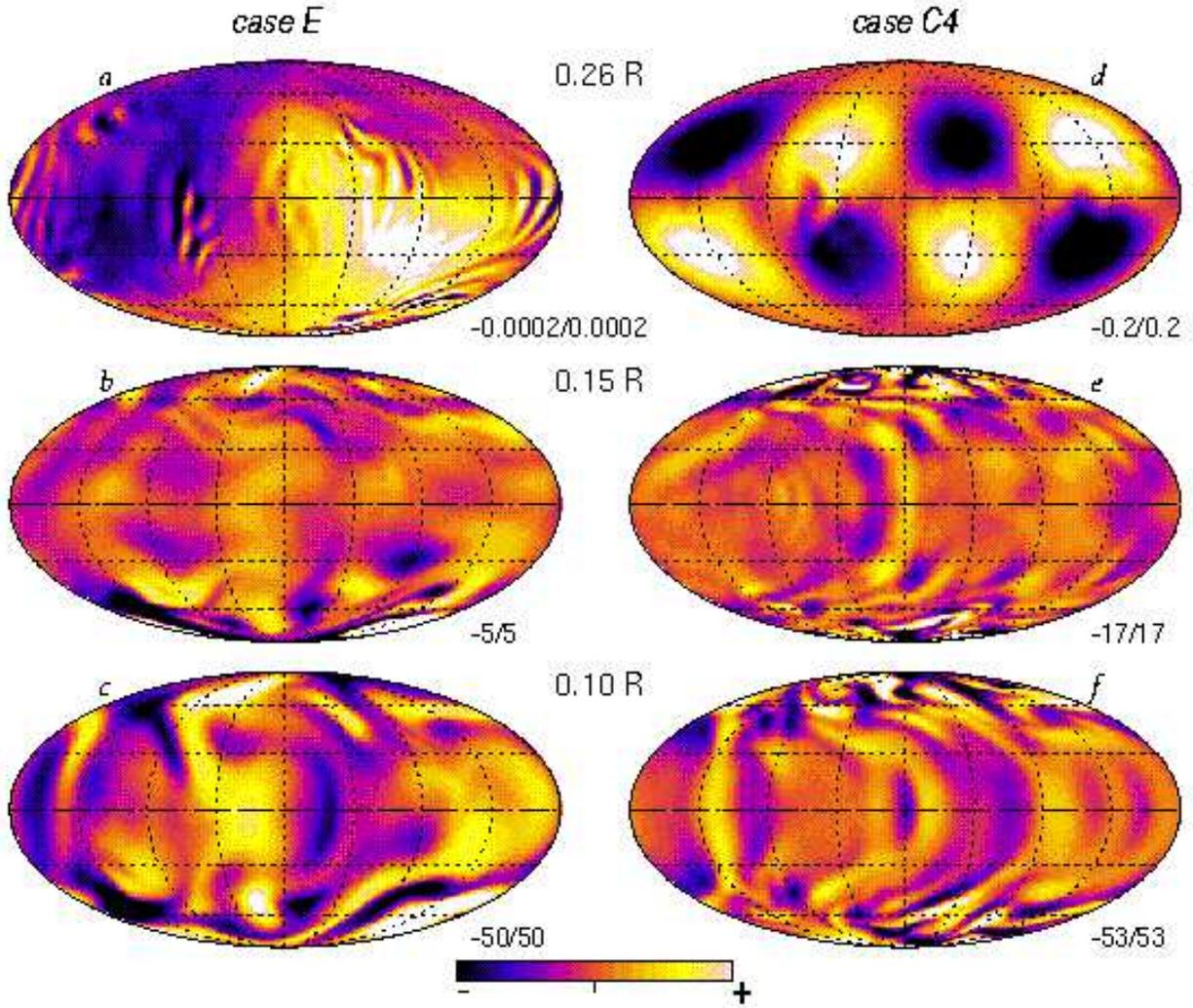}
  \caption{The excitation of gravity waves in the radiative envelope for
  cases E (left) and C4 (right), shown in global mappings of radial
  velocity $v_r$ at one instant in time (top panels, sampling $r=0.26R$).
  The associated convective flows are also shown within the overshoot
  region (middle panels, $r=0.15R$) and in the convective core (lower
  panels, $r=0.10R$).  For case E, gravity waves of small amplitude
  (scaling indicated) are visible in ($a$) as weak ripples in the largely
  quiescent radiative zone.  For case C4, the waves appear in ($d$) as a
  prominent global resonance of greater amplitude.}
\end{figure}
	
\begin{figure}[hpt]
  \center
  \epsscale{1.0}
  \plotone{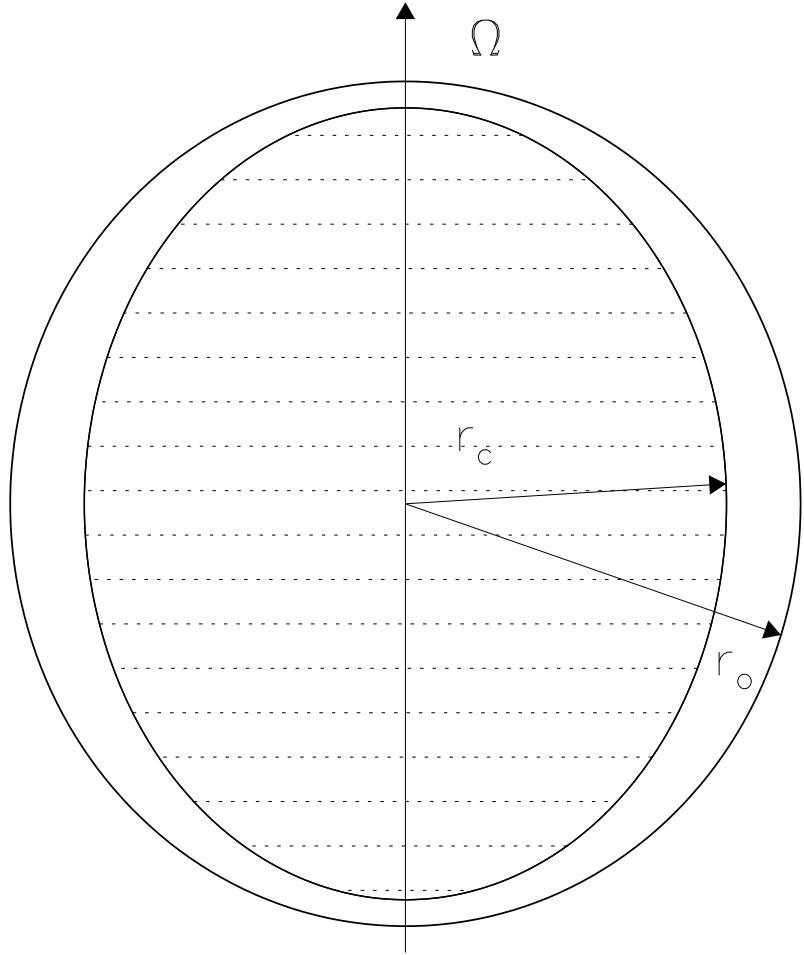}
  \caption{Schematic of the typical geometry of the prolate convective core
  of radius $r_{c}(\theta)$, and of the further region of overshooting,
  which extends to radius $r_o$ largely independent of latitude
  $\theta$. The nearly adiabatic convective interior has greatest spatial
  extent along the axis of rotation (here vertical).}
\end{figure}

\begin{figure}[hpt]
  \center
  \epsscale{1.0}
  \plotone{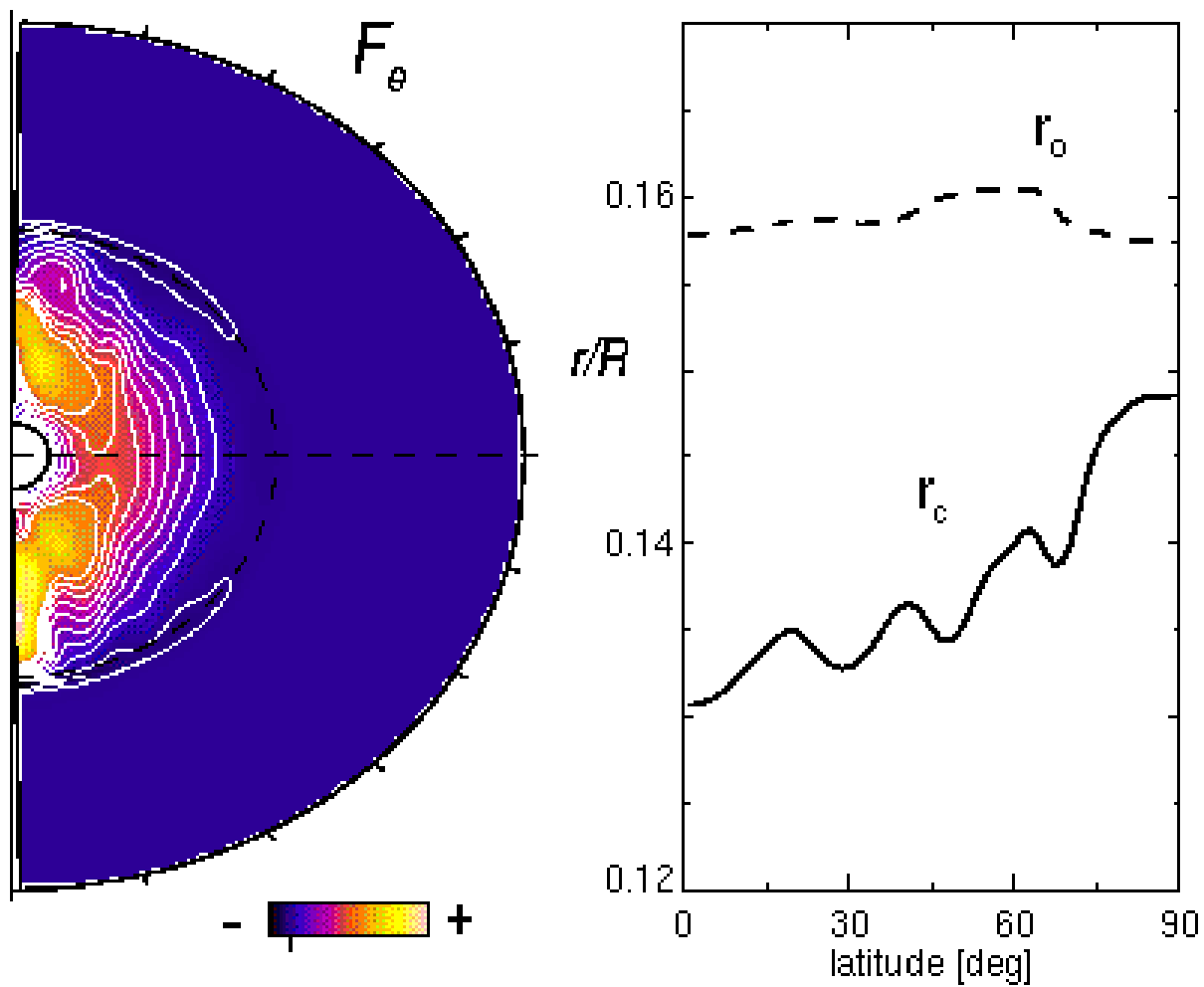}
  \caption{($a$) Variation with radius and latitude of the radial enthalpy
  flux $\hat{F_e}$ in case C, averaged over 180 days.  This reveals the complex patterns to the
  transport achieved by the vigorous convection within the core (with
  positive flux denoted in yellow/red tones), and the weaker response in
  the overshooting region (with a negative flux).  The equator is denoted
  by the dashed line. ($b$) Assessment of $r_c$ and $r_o$ for the same
  case, expressed in proportional radius as a function of latitude.  Using
  $\hat{F_e}$, $r_c$ (solid line) measures the radius at the outer edge of
  the convective core at which the enthalpy flux becomes negative, and
  $r_o$ the radius of overshooting at which that flux essentially
  vanishes.}
\end{figure}

\begin{figure}[hpt]
  \center
  \epsscale{1.0}
  \plotone{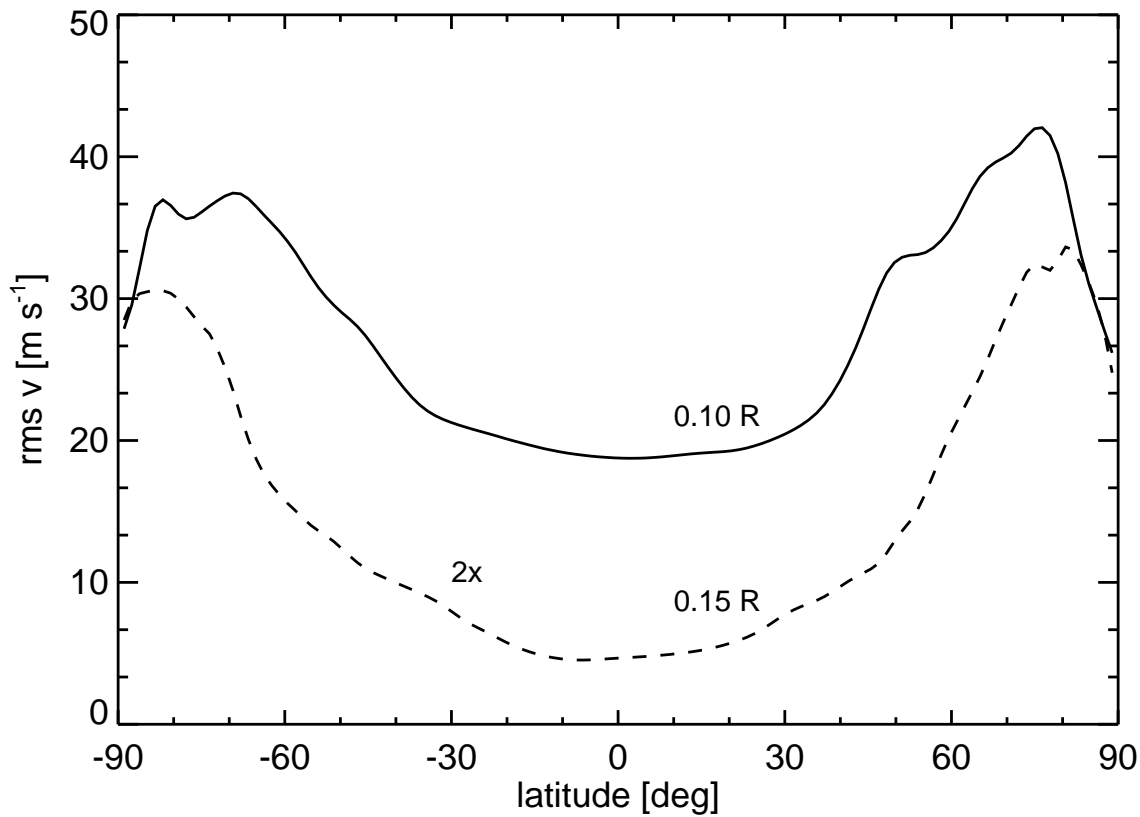}
  \caption{Variation with latitude of the root-mean-square of radial
  velocities $\vrr$ in case C within both the convective core
  ($r=0.10R$) and the region of overshooting ($r=0.15R$), averaged over 180
  days. The latter
  velocities have been multiplied by a factor of two for visibility.  Radial velocities are
  highest near the poles, and lowest near the equator, owing partly to the
  deflection of outward motions by the Coriolis forces in these rotating
  domains.  The prolate geometry of the convective core appears to arise from
  such latitudinal variation of $v_r$.}
\end{figure}

\begin{figure}[hpt]
  \center
  \epsscale{0.71}
  \plotone{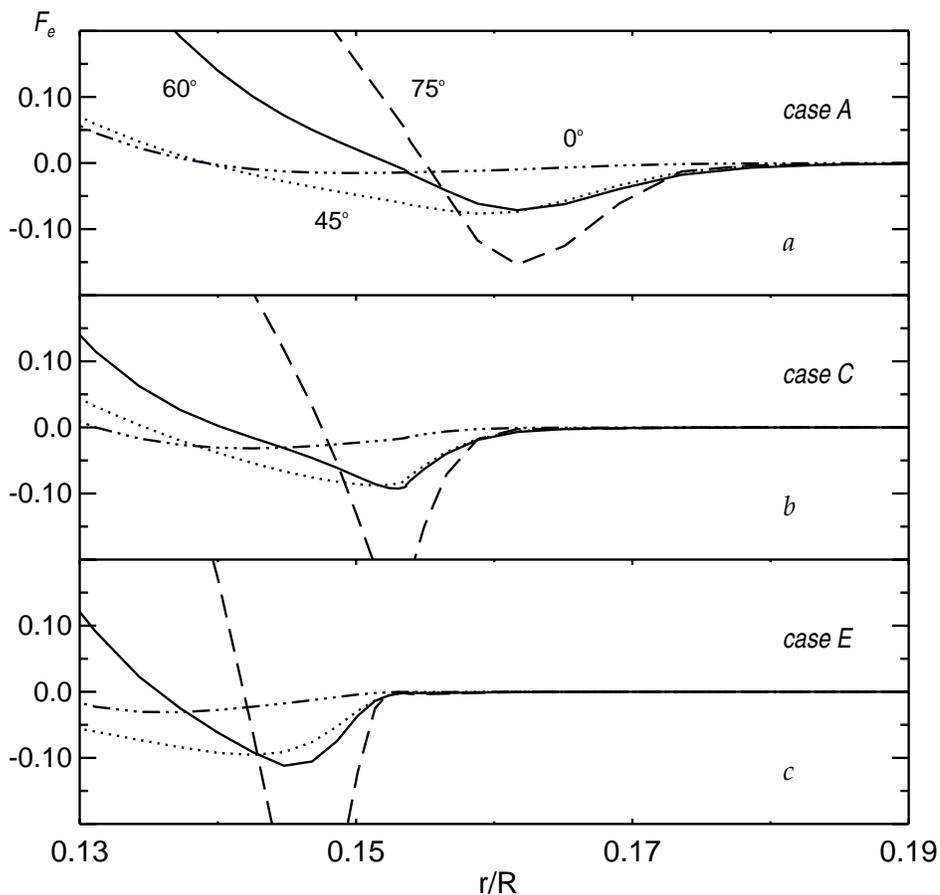}
  \vspace{0.5in}
  \caption{Variation of radial enthalpy flux $\hat{F_e}$ with radius at
  selected indicated latitudes for cases A, C, and E involving
  progressively stiffer stable stratifications in the radiative exterior.
  The radius $r_c$ at which the enthalpy flux first becomes negative
  (providing an estimate of the boundary of the convective core) is a
  strong function of latitude, with higher latitudes having a core of
  greater spatial extent.  In contrast, the radius $r_o$ at which the
  negative enthalpy flux in the overshooting region essentially vanishes is
  largely insensitive to latitude.  Increasing stiffness decreases the
  radii at which these transitions occur.}
\end{figure}

\begin{figure}[hpt]
  \center
  \epsscale{0.55}
  \plotone{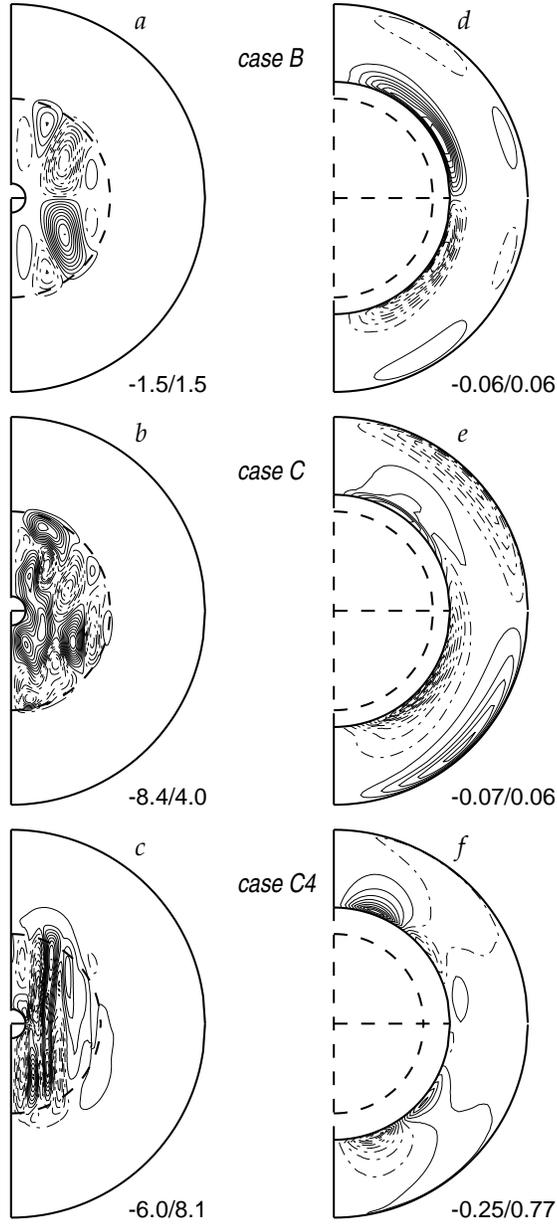}
  \caption{Time-averaged streamfunctions of meridional mass flux for cases
  B (upper panels), C (middle), and C4 (lower), displayed for the whole
  domain (left column) and at much amplified scaling for the outer portion
  of the radiative zone only (right).  The time intervals used in the
  averaging are all of order 110 days.  Solid lines denote counter-clockwise
  flow, dashed lines clockwise.  The associated maximum and minimum velocities for
  $\vtr$ (in m s$^{-1}$) are indicated.  }
\end{figure}

\begin{figure}[hpt]
  \center
  \epsscale{1.0}
  \plotone{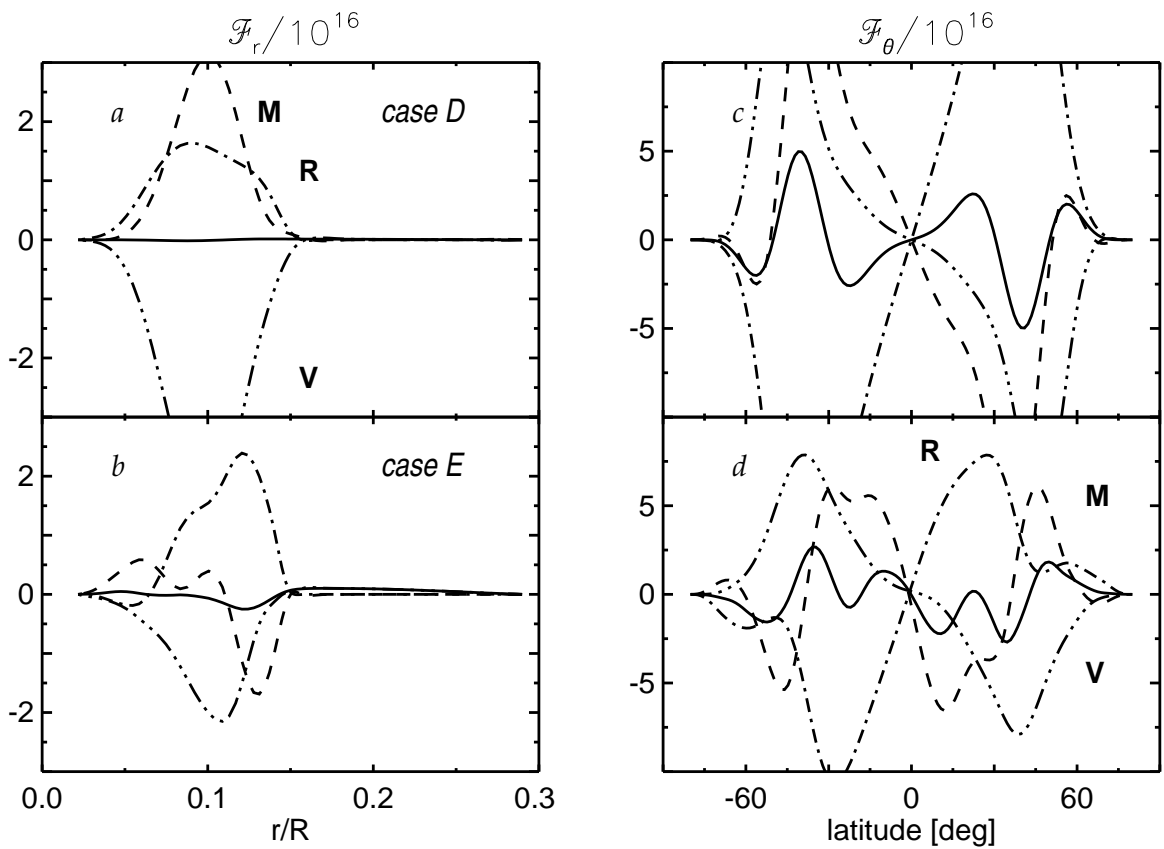}
  \caption{Time average of the latitudinal integral of the angular
  momentum flux ${\cal F}_r$ (left panels) and of the radial integral
  of the angular momentum flux ${\cal F}_{\theta}$ (right panels) for cases
  D (upper row) and E (lower row).  The time intervals used in forming the
  averages were 670 days for case D and 620 days for case E.  The fluxes have been decomposed into
  their viscous (V), Reynolds stress (R), and meridional circulation (M)
  components, and the solid curves represent the total fluxes.  Positive
  quantities represent fluxes radially outward, or latitudinally from north to south.}
\end{figure}

\begin{figure}[hpt]
  \setlength{\unitlength}{1.0cm}
\begin{picture}(5,5.2)
\includegraphics{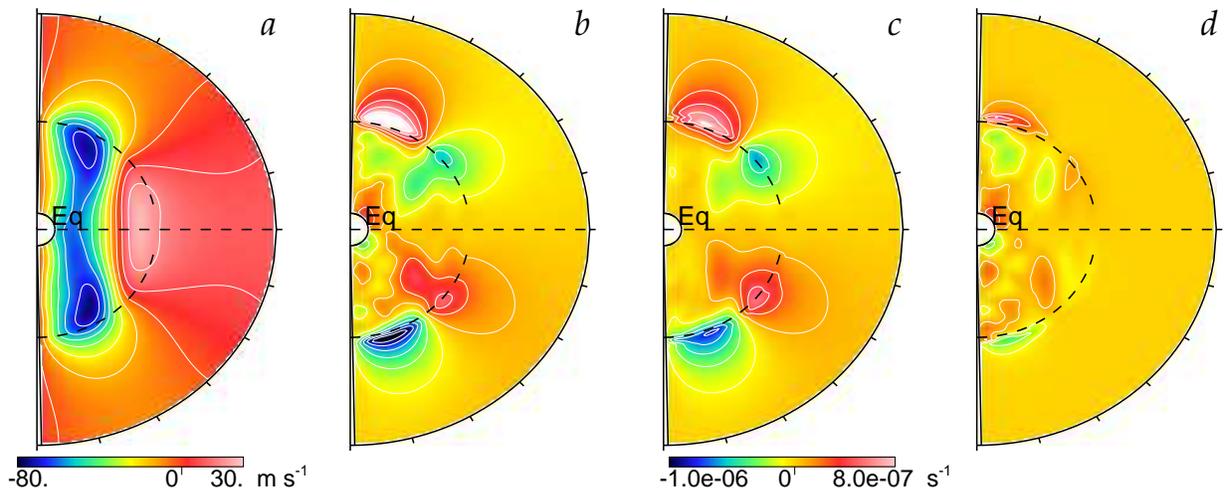}
\end{picture}
  \caption{Temporal (640-day) and longitudinal averages for case E showing ($a$) the
  longitudinal velocity $\vph$, ($b$) $d\vph/dz$, ($c$) the baroclinic term
  in the meridional force balance, and ($d$) the difference between the
  latter two.  The same color scale is used in ($b$), ($c$), and
  ($d$). Near the interface between the convective core and the radiative
  envelope, baroclinic effects account for much of the differential
  rotation.  Within the bulk of the prolate convective core, Reynolds and
  viscous stresses have a significant role in driving those zonal flows.  }
\end{figure}

\end{document}